\documentclass{aa}  
\newcommand{\sub}[1]{_{\rm #1}}
\newcommand{\CII}{[C\,{\sc ii}]}   
\newcommand{\CI}{[C\,{\sc i}]}  
\newcommand{\thirteenCII}{[$^{13}$C\,{\sc ii}]}
 
\newcommand{\OI}{[O\,{\sc i}]}
\newcommand{\OIII}{[O\,{\sc iii}]} 

\newcommand{\NIII}{[N\,{\sc iii}]}   
\newcommand{\HII}{H\,{\sc ii}}   
\newcommand{\HI}{H\,{\sc i}}   
   
\newcommand{\emm}[1]{\ensuremath{#1}}   
\newcommand{\emr}[1]{\emm{\mathrm{#1}}} 
\newcommand{\unit}[1]{\emm{\, \emr{#1}}}

\newcommand{\pscm}{\unit{cm^{-2}}}
\newcommand{\ccm}{\unit{cm^{3}}}
\newcommand{\pccm}{\unit{cm^{-3}}}
\newcommand{\kms}{\unit{km\,s^{-1}}}
\newcommand{\micron} {\unit{\mu m}}

\newcommand{\chr}{^{\rm h}}
\newcommand{\cmin}{^{\rm m}}
\newcommand{\csec}{^{\rm s}}
\newcommand{\degree}{^{\circ}}
\newcommand{\changed}{}
\usepackage{graphicx}
\usepackage[english]{babel}
\usepackage{txfonts}
\usepackage{natbib}
\begin{document}

   \title{The fine structure line deficit in S~140}

   \author{V. Ossenkopf\inst{1}
          \and
          E. Koumpia\inst{2,3}
          \and
          Y. Okada\inst{1}
          \and
          B. Mookerjea\inst{4}
          \and
          F.F.S. van der Tak\inst{2,3}
          \and
          R. Simon\inst{1}
          \and
          P. P\"utz\inst{1}
          \and
          R. G\"usten\inst{5}
	  }

   \institute{I. Physikalisches Institut der Universit\"{a}t zu K\"{o}ln, 
	Z\"{u}lpicher Stra\ss{}e 77, 50937, K\"{o}ln, Germany\\ \email{ossk@ph1.uni-koeln.de}
	\and
	SRON Netherlands Institute for Space Research, Landleven 12, 9747 AD Groningen, 
	The Netherlands; 
	\and
	Kapteyn Institute, University of Groningen, PO Box 800, 9700 AV Groningen, 
  	The Netherlands
	\and
	Department of Astronomy and Astrophysics, Tata Institute of Fundamental Research, Homi Bhabha Road, Colaba, 400005, Mumbai, India
	\and
	Max-Planck-Institut für Radioastronomie, Auf dem H\"ugel 69, 53121 Bonn, Germany
        }

   \date{Received ; accepted}

 
  \abstract
   {}
   {We try to understand the gas heating and cooling in the S~140 star 
   forming region by spatially and spectrally resolving the distribution
   of the main cooling lines with \textit{GREAT/SOFIA} and combining our data with
   existing ground-based and \textit{Herschel} observations that trace the
   energy input and the density and temperature structure of the source.
   }
   {We mapped the fine structure lines of \OI{} (63~\micron{}) and \CII{} 
    {(\changed 158~\micron{})} and 
    the rotational transitions of CO 13--12 and 16--15 with 
    GREAT/SOFIA and analyzed the spatial
    and velocity structure to assign the emission to individual heating
    sources. We measure the optical depth of the \CII{} line and
    perform radiative transfer computations for all observed transitions.
    By comparing the line intensities with the far-infrared continuum
    we can assess the total cooling budget and measure the gas
    heating efficiency.
   }
    {The main emission of fine structure lines in S~140 stems from
    a 8.3$''$ region close to the infrared source IRS~2
    that is not prominent at any other wavelength. It can be
    explained by a photon-dominated region (PDR) structure around
    the embedded cluster if we assume that the \OI{} line intensity
    is reduced by a factor seven due to self-absorption. The external
    cloud interface forms a second PDR at an inclination of 80--85
    degrees illuminated by an UV field of 60 times 
    the standard interstellar radiation field. The main radiation 
    source in the cloud, IRS~1, is not prominent at all in the
    fine structure lines. We measure line-to-continuum cooling
    ratios below $10^{-4}$, i.e. values lower than in any other
    Galactic source, rather matching the far-IR line deficit seen 
    in ULIRGs. In particular the low intensity of the \CII{} line
    can only be modelled by an extreme excitation gradient in
    the gas around IRS~1. We found no explanation why IRS~1 shows no
    associated fine-structure line peak, while IRS~2 does.}
   {The inner part of S~140 mimics the far-IR line deficit 
    in ULIRGs thereby providing a template that may lead to a future model.}


\authorrunning{Ossenkopf et al.} 
\titlerunning{Fine structure line deficit in S~140}
   \maketitle
%

\section{Introduction}

The cooling budget of the gas in galaxies is dominated by the emission in
the fine structure lines of atomic oxygen and ionized carbon 
\citep[see e.g.][]{Burton1990, Roellig2007}. Their emission stems to a large
fraction from UV-illuminated surfaces of molecular clouds, so called
photon-dominated regions \citep[PDRs][]{TielensHollenbach1985,SternbergDalgarno1995}.
Due to the clumpy or fractal structure of molecular clouds, the PDR
surfaces may represent a large fraction of their total volume 
\citep{Ossenkopf2007}. To quantify the importance of the PDRs in terms of the overall
cooling balance of molecular clouds, we have to compare the
fine structure line emission with the lines of CO (and its isotopologues)
that measure the emission of the molecular material and to the continuum
emission from interstellar dust. The dust emission measures the
combination of the total column density structure with the local heating, 
optically thin molecular lines provide us with a view on the spatial and
velocity distribution of the molecular gas, and the fine structure lines
contain the information on the local energy input through UV radiation
and the cooling balance in the gas. The German REceiver 
for Astronomy at Terahertz Frequencies (GREAT) onboard the Stratospheric 
Observatory for Infrared Astronomy (SOFIA)
allows for the first time to obtain velocity resolved
spectra of the 63~\micron{} line of atomic oxygen, \OI{}, with 
a good sensitivity {\changed \citep{Buechel2015}}.
We used the instrument to measure the line simultaneously with the
158~\micron{} line of ionized carbon, \CII{}, in the S~140 star-forming 
region\footnote{Comparison to integrated line observations from the Herschel 
Space Observatory are provided in 
Appx.~\ref{sect_appx_herschel}.\\{\it Herschel} is 
an ESA space observatory with science instruments provided by European-led 
Principal Investigator consortia and with important participation from NASA.}.

S~140 is an \HII{}-region at the surface of the L~1204 molecular cloud,
created by the illumination from the B0V star HD~211880, at a distance 
of 764~pc \citep{Hirota2008}. The external surface of the molecular cloud
is exposed to a moderate UV field of $140\,\chi_0$ \citep[][]{PoelmanSpaans2005}
to $230\,\chi_0$ \citep{Timmermann1996} \citep[$\chi_0$ =
standard interstellar radiation field][]{Draine1978}. 
However, ongoing star-formation within the cloud already formed 
a cluster of deeply embedded radiation sources \citep{Evans1989}, 
probably creating more internal PDRs around  the sources. Radio continuum
observations detected ultracompact \HII{} regions at the strongest
sources \citep[e.g.][]{Tofani1995, Hoare2006}. The strongest submm source,
IRS~1, harbors a cluster of massive young stellar objects at a projected
distance of 75$''$ from the cloud interface \citep{Minchin1993}. 
\citet{Weigelt2002} showed the presence of an outflow cavity around IRS1. The 
corresponding outflow walls are illuminated by the cluster so that they
form internal high-density PDRs. \citet{Dedes2010} suggested that
the \CII{} emission from IRS1 measured by HIFI onboard Herschel stems from 
these irradiated outflow walls, exposed to much higher radiation fields
of more than $10^5\,\chi_0$. Physical models of the structure around IRS~1 
have been proposed by \citet{Harvey1978}, \citet{Guertler1991},
\citet{Minchin1993}, \citet{vanderTak2000}, \citet{Mueller2002},
\citet{deWit2009}, \citet{Maud2013}. The whole cloud is fragmented and
clumpy. \citet{PoelmanSpaans2006} determined
densities of n$\sub{H_2}\approx 4\times 10^{5}$~cm$^{-3}$ for the clumps
and n$\sub{H_2}\approx 10^{4}$~cm$^{-3}$ for the interclump medium.

In Sect. 2, we present the new GREAT observations of the 63~\micron{} \OI{}
and 158~\micron{} \CII{} fine structure and the CO 16--15 and 13--12 
lines towards S~140. In Sect. 3, we discuss the morphology and velocity
structure of the observed emission compared to other lines, in particular
low-$J$ CO transitions, and literature data. In Sect. 4, we derive gas
parameters across the map and for selected positions to understand 
the nature of the emitting gas. Sect.~5 discusses possible reasons for 
the association of the fine structure emission peak with an otherwise
inconspicuous source and the overall extremely low fine-structure
line emission.
\begin{table*}[t]
\caption{Parameters of the GREAT line observations.}
\centering
\begin{tabular}{l l c | c c | c c c c c}
\hline
\hline
Species & Transition & $\nu_0$ & E$_{up}$ & $n\sub{crit}$\tablefootmark{a} & beam FWHM 
& $\eta\sub{mb}$ & $\Delta v$ & $\sigma_{i,\rm rms}$\tablefootmark{b} & $\sigma\sub{map,rms}$\tablefootmark{c}\\
 &  & [MHz] & [K] & [cm$^{-3}$] & [arcsec] & & [\kms{}] & [K] & [K] \\
\hline
C$^+$ & \CII{} $^2P_{3/2}-^2P_{1/2}$ & 1900536.90 & 91.21 & $3\times 10^3$ 
& 14.1 & 0.67 & 0.3 & 3.7 & 0.8-1.2 \\
O & \OI{} $^3P_1-^3P_2$ & 4744777.49 & 227.71 & $5\times 10^5$ 
& 6.6 & 0.67 & 0.4 & 3.2 & 1.3-1.9 \\
CO & 13--12 & 1496922.91 & 503.10 & $1\times 10^6$ & 
18.6 & 0.70 & 0.3 & 1.9 & 0.9-1.3\\
   & 16--15 & 1841345.51 & 751.70 & $2\times 10^6$ & 
     14.5 & 0.69 & 0.4 & 2.9 & 0.7-1.0\\
\hline
\label{tab_greatlines}
\end{tabular}
\tablefoot{
\tablefoottext{a}{Collision rates from {\changed
\citet[][at 50K]{WiesenfeldGoldsmith2014,Jaquet1992,Yang2010}}.
For optically thick lines the effective critical 
densities can be considerably lower due to radiative trapping.}
\tablefoottext{b}{Noise in individual spectra.}
\tablefoottext{c}{Noise in Nyquist-sampled map with GREAT beam FWHM.}
}
\end{table*}

\section{GREAT observations}

We used the German REceiver for Astronomy at Terahertz Frequencies
\citep[GREAT\footnote{GREAT is a development by the MPI für 
Radioastronomie and the KOSMA / Universit\"at zu K\"oln, in
cooperation with the MPI für Sonnensystemforschung and the DLR 
Institut für Planetenforschung},][]{GREAT} onboard the Stratospheric Observatory
for Infrared Astronomy \citep[SOFIA,][]{SOFIA} to take on-the-fly
maps of the central region of S~140 in \OI{}, 63\micron{},
\CII{}, CO 16--15 and CO 13--12. 

The \OI{} observations were performed in the nights from
May 16 and 17, 2014. The
\CII{} data were taken on May 16, 17, and 20, 2014. CO 13-12
was observed in two outer strips on May 20, 2014 and CO 16-15
was measured in the inner strip on January 13, 2015.

We mapped an area of $144 \times 96''$ oriented at an angle of 37$\degree$ 
east of
north, perpendicular to the outer interface of the PDR. Due to 
instrument scheduling constraints, the \OI{} and CO 16-15 observations 
were restricted to an inner strip of only 40$''$ width covering the
central cluster, while the observation of the CO 13-12 line avoided
the central 32$''$ covering only the outer two strips. In \CII{} we
have the complete map covering the whole area. All observations involving
\OI{} were taken on a 3$''$ grid to obtain full sampling, the outer
strips were mapped with an 8$''$ sampling, guaranteeing a close-to full
sampling for the telescope beam at the \CII{} frequency. The OFF position
for the observations was located south-west of \HII{} region and molecular
cloud at R.A.$=22\chr{}18\cmin{}37\csec{}.0$, Dec$=63\degree{}14'18''.0$ (J2000), six arcminutes away from IRS~1, our reference position in the 
maps at R.A.$=22\chr{}19\cmin{}18\csec{}.21$, Dec$=63\degree{}18'46''.9$.
XFFT spectrometers with a resolution of 44~kHz were used as backends.

Atmospheric calibration was done through the standard GREAT pipeline 
\citep{Guan2012}. The pointing accuracy of the telescope and the 
alignment of the  GREAT instrument relative to the telescope optical axis
give a combined pointing error of less than 2~arcseconds. The two GREAT
channels operating in parallel were co-aligned to 1~arcsec accuracy. The forward
efficiency is 0.97 for all observed frequencies. The other parameters
of the GREAT beam at the observed frequencies are given in 
Tab.~\ref{tab_greatlines}. All spectra were scaled to main beam 
Rayleigh-Jeans temperatures.

For the spectra of CO and \CII{} the default pipeline provided quite
flat baselines. To be independent of the continuum level, a first order
baseline, measured over 100~\kms{} outside of the 15~\kms{} of the line
was subtracted. As the \OI{} maps suffered
from some OTF-striping, these data were further processed by a secondary 
OFF subtraction, treating the first and last points of all OTF lines as 
reference allowing for the correction of gain drifts. This assumes that
the map edges fall in regions free of emission, an assumption that is
confirmed when inspecting the raw data. To improve the signal
to noise ratio, the spectra were smoothed with a box-car filter to a
resolution of 0.3--0.4~\kms{}. 
Due to the different combinations of observations, the total noise
in the sum spectra taken within one beam is variable over the map.
The average rms noise levels of the individual spectra and the 
beam-averaged sum spectra across the map at the smoothed resolution
are given in Tab.~\ref{tab_greatlines}.

Some of the \OI{} spectra showed
more irregular, wavy baseline structures not described by a simple
linear relation. We tried to correct them by fitting various orders
of polynomials to the baseline varying the order from two to ten. 
We did not find a significant global improvement of the spectra with
increasing order, but an improvement of a few individual spectra.
By using the criterion of the least negative values in the integrated
intensity map, we selected the seventh order for all spectra
but use the outcome of the variation of the baseline orders
basically as an error estimate for the baseline uncertainty. 
For different baseline orders, the intensities of
individual bins in the line changed by up to 0.4~K and the 
integral over the full velocity range of the line ($-15 - 0$~\kms{}) 
varied by up to 6~K~\kms{}.
This uncertainty has to be added to the noise and the calibration error 
of the instrument of approximately 20\,\% as the uncertainty of the data.

\section{Spatial structure of the emission}

   \begin{figure}
   \includegraphics[angle=90,width=7.8cm]{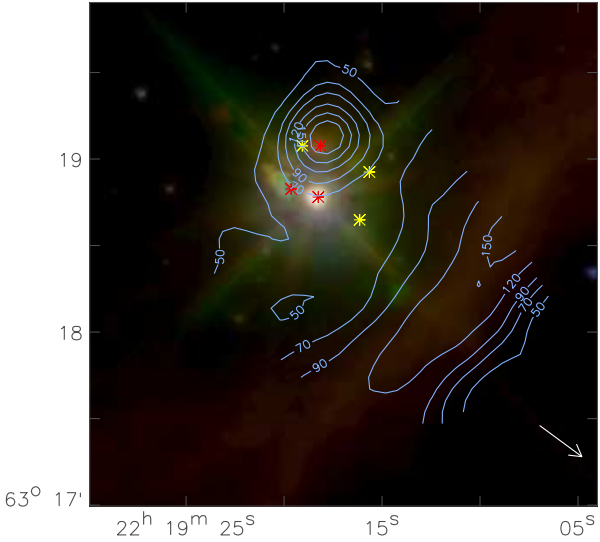}
      \caption{Integrated intensity map of \CII{} overlaid on a Spitzer IRAC
		false-color image of the same region. The IRAC map 
		uses the logarithmic intensities of the 3.6, 5.6 
		and 8\micron{} channels. The 
		integrated intensity contours are drawn at 
		50, 70, 90, 120, 150, 180 K \kms{}. 
		The external cloud interface is visible in the south-west.
		{\changed The arrow shows the direction towards the 
		illuminating source HD~211880.} The red marks
		indicate the position of the known infrared sources: 
        south-west is IRS~1, the northern one is IRS~2, and 
        the south-east we find IRS~3 \citep{Evans1989}, 
        the yellow marks show position
		of known submm cores \citep{Minchin1995,Maud2013}. 
		The IRAC peak falls at the location of IRS~1.}
         \label{fig_cii-spitzer}
   \end{figure}
   \begin{figure}
   \centering
   \includegraphics[angle=90,width=\hsize]{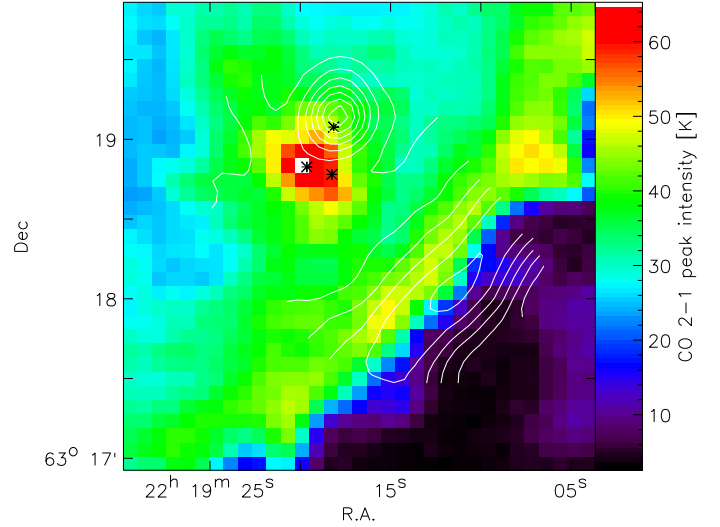}
      \caption{Peak intensity map of \CII{} overlaid on a IRAM
		30m map of the CO 2--1 peak intensity. The \CII{} contours
 		are drawn at 15, 20, 25, 30, 35, 40, 45~K. The black marks
		indicate the positions of IRS~1-3.}
         \label{fig_cii-peakintens}
   \end{figure}

Figure~\ref{fig_cii-spitzer} shows the observed integrated intensity
map of the \CII{} line superimposed on a Spitzer IRAC
image of the cloud composed of the 3.6, 5.6 and 8\micron{} channels.
The IRAC image emphasizes two structures in the cloud.
The peak in all bands is given by the embedded source IRS~1,
relatively deep in the cloud. The embedded cluster produces
about $10^4$ solar luminosities, $L_\odot$, heating the dust to 
temperatures of up to 1400~K \citep{Koumpia2015}. In the south-west 
we see the cloud surface illuminated by HD~211880 as a red band-like
structure due to the excitation of PAHs at this external PDR. On top
of these dominant features, we see some smaller spots east of IRS~1
close to IRS~3, bright in 5.6~\micron{} emission. The \CII{} 
emission traces the front of the PDR in the south-west {\changed
through a secondary peak}, but has
a clear maximum 20$''$ north of IRS~1, slightly north of IRS~2.

To compare the fine structure lines with the molecular cloud material
we use the maps of CO 1--0, 2--1, C$^{18}$O 1--0, and $^{13}$CO
1--0 taken with the IRAM 30m telescope, presented and discussed by
\citet{Koumpia2015}. Figure~\ref{fig_cii-peakintens} compares the 
\CII{} peak intensities with the CO 2--1 peak intensity map. 
{\changed The spatial distribution of the CO 2--1 line shows a 
behavior similar to the mid-infrared continuum. CO 2--1 peaks 
about 20$''$ south of the \CII{} peak in a region covering IRS~1
and IRS~3. The CO 2--1 maximum falls close to IRS~3 and at the cloud surface 
the line is brighter than deep in the cloud. There } is a clear
layering with \CII{} peaking closer to the illuminating star
and CO 2--1 deeper in the molecular cloud, as expected from standard
PDR models \citep[see e.g.][]{HollenbachTielens1999}. The location of the 
peak of \CII{} emission at the cloud surface therefore represents an expected
morphology, but the relative shift of the emission peak to the north of 
IRS~1 is a striking unexpected feature in the \CII{} maps.

   \begin{figure}
   \centering
   \includegraphics[angle=90,width=\hsize]{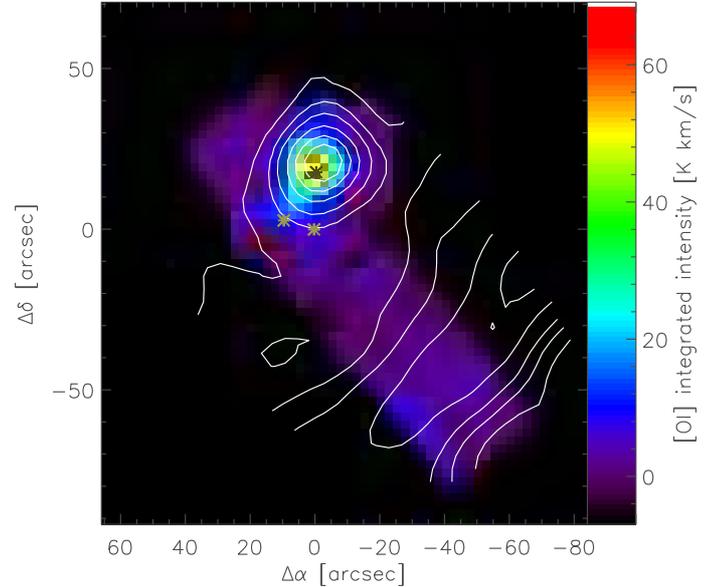}
      \caption{Integrated intensity map of the \OI{} 63~\micron{}
		line (colors) overlaid by contours of the \CII{} 
		integrated intensity (levels as in Fig.~\ref{fig_cii-spitzer}).
		The black marks indicate the positions of IRS~1-3.
        All coordinates are taken relative to the position of 
        IRS~1 at R.A.$=22\chr{}19\cmin{}18\csec{}.21$, Dec$=63\degree{}18'46''.9$ (J2000)}
         \label{fig_oi-cii-integrated}
   \end{figure}

Figure~\ref{fig_oi-cii-integrated} compares the \CII{} with
the \OI{} 63~\micron{} line integrated intensity. The \OI{} map shows 
a good match of the emission peak with that of \CII{}.
The peak is strongly concentrated and dominates the overall map.
\OI{} also peaks slightly north of IRS~2; it may be offset from 
the center of the \CII{} peak by about 2$''$, but this is close
to our pointing accuracy. 
The location of the \CII{} and \OI{} peak indicates an association
with IRS~2, similar to an already known small continuum feature 
visible at mid-infrared wavelengths \citep{Harvey2012}.
The \OI{} peak is slightly elongated towards the south-east 
in the direction towards IRS~3. IRS~1 is not prominent at all in
\OI{}. We also see a slightly increased \OI{} intensity at the cloud 
interface in the region where \CII{} peaks. Unfortunately, the
data quality does not allow to discriminate whether there is any
relative offset between the two fine structure line peaks in that
region.

   \begin{figure}
   \centering
   \includegraphics[angle=90,width=\hsize]{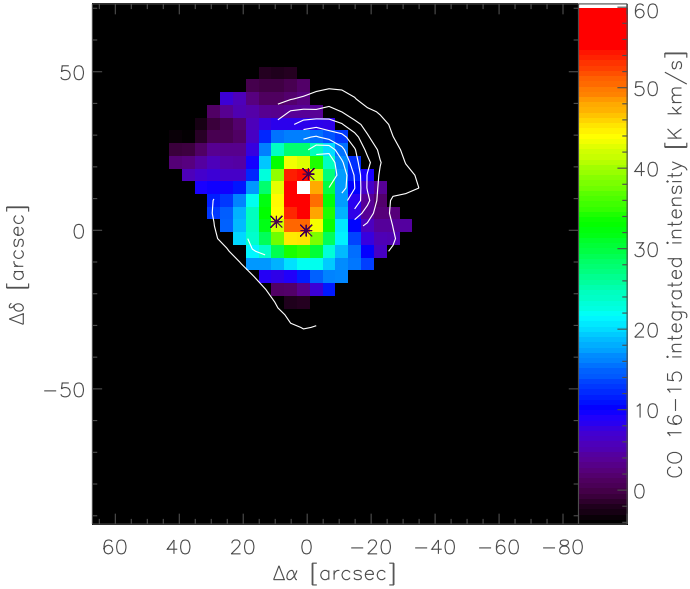}
      \caption{Integrated intensity map of CO 16--15 with contours
        of CO 13--12 drawn at levels of 10, 20, 30, 40, 50, 60, 
        70~K~\kms{}. Only the region around IRS~1 has been observed 
        in CO 16--15; in CO 13--12 only two outer strips have been
        covered avoiding the center. The black marks
		indicate the positions of IRS~1-3.}
         \label{fig_co-lines-integintens}
   \end{figure}
   
Figure~\ref{fig_co-lines-integintens} shows the integrated
intensities of the CO 13--12 and 16--15 lines. Unfortunately,
only complementary regions were mapped in the two lines with
narrow overlapping strips. Therefore, the CO 13--12 
observations are hardly useful to understand the nature of the 
peak seen in the fine-structure lines. We see, however, that
both high-$J$ CO lines show a more extended emission
around the central sources compared to the fine-structure lines.
The CO 16--15 peak falls between IRS~1 and IRS~2, i.e. we see a
superposition of hot CO emission from both IRS~1 and from the 
fine-structure line peak close to IRS~2. The high-$J$
CO lines trace both components in a similar way; they show an 
intermediate behavior between the fine-structure lines and the
low-$J$ CO.

\subsection{Line profiles}
\label{sect_line_profiles}

   \begin{figure*}
   \centering
   \includegraphics[angle=90,width=6cm]{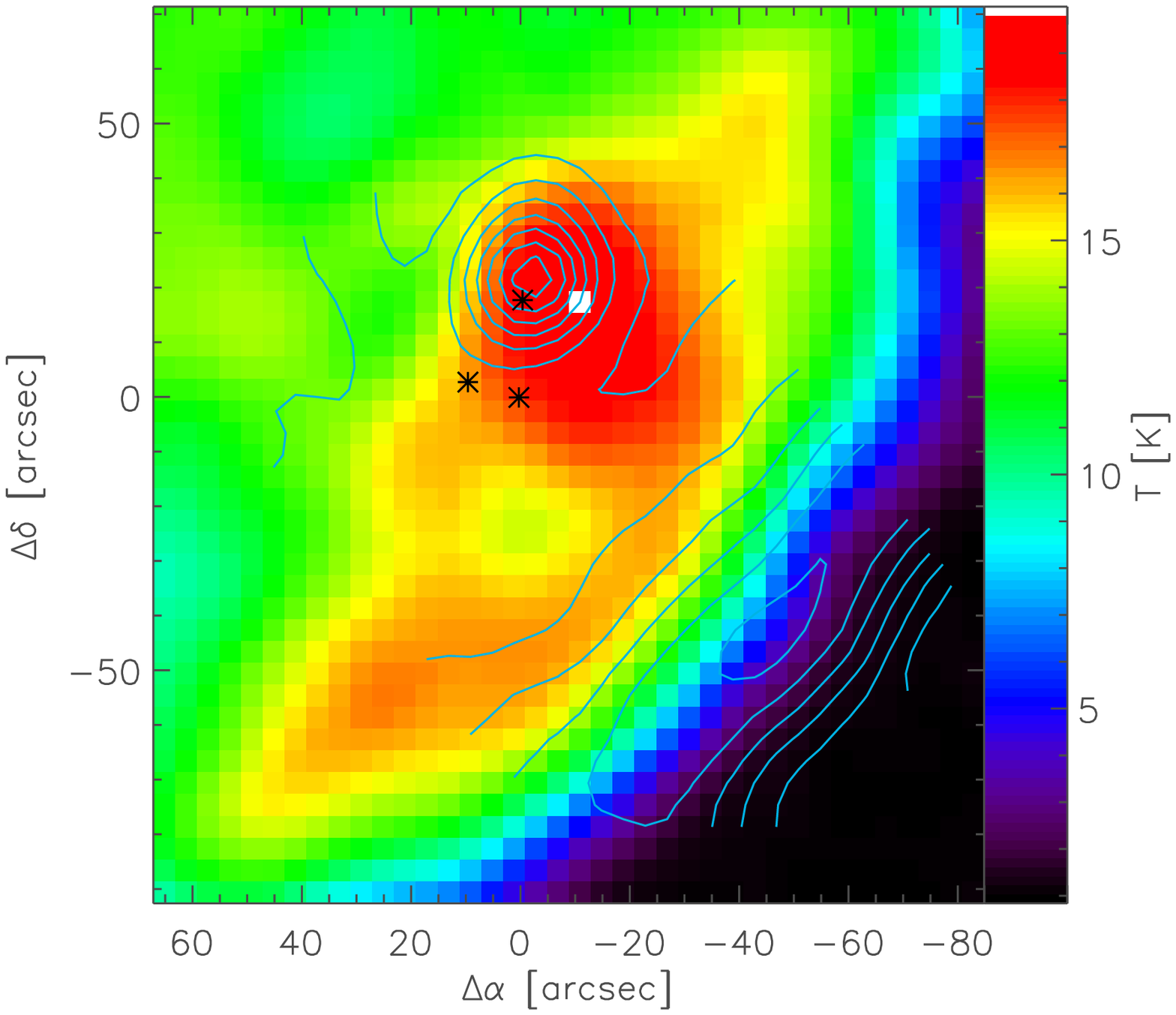}
   \includegraphics[angle=90,width=6.1cm]{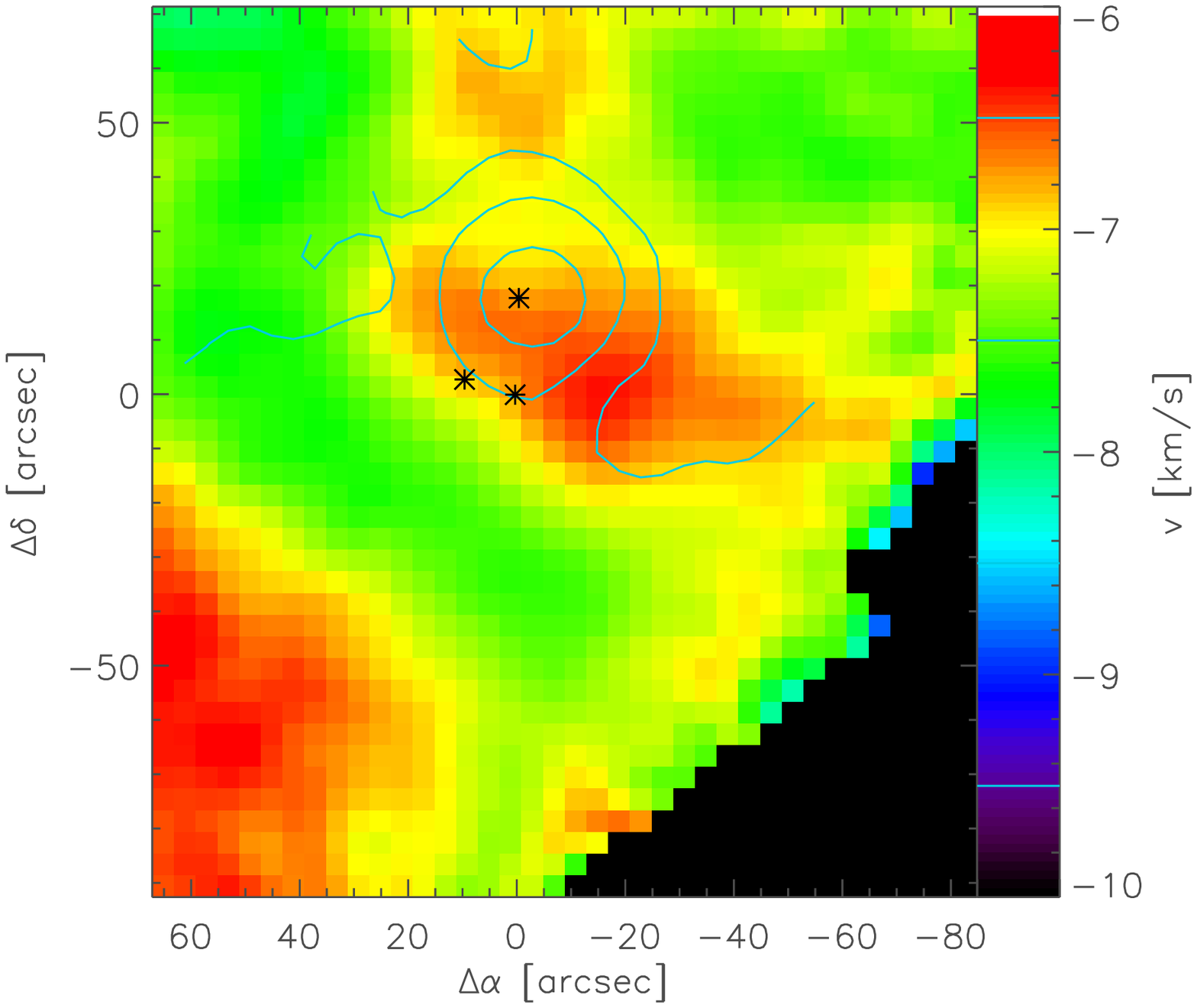}
   \includegraphics[angle=90,width=5.8cm]{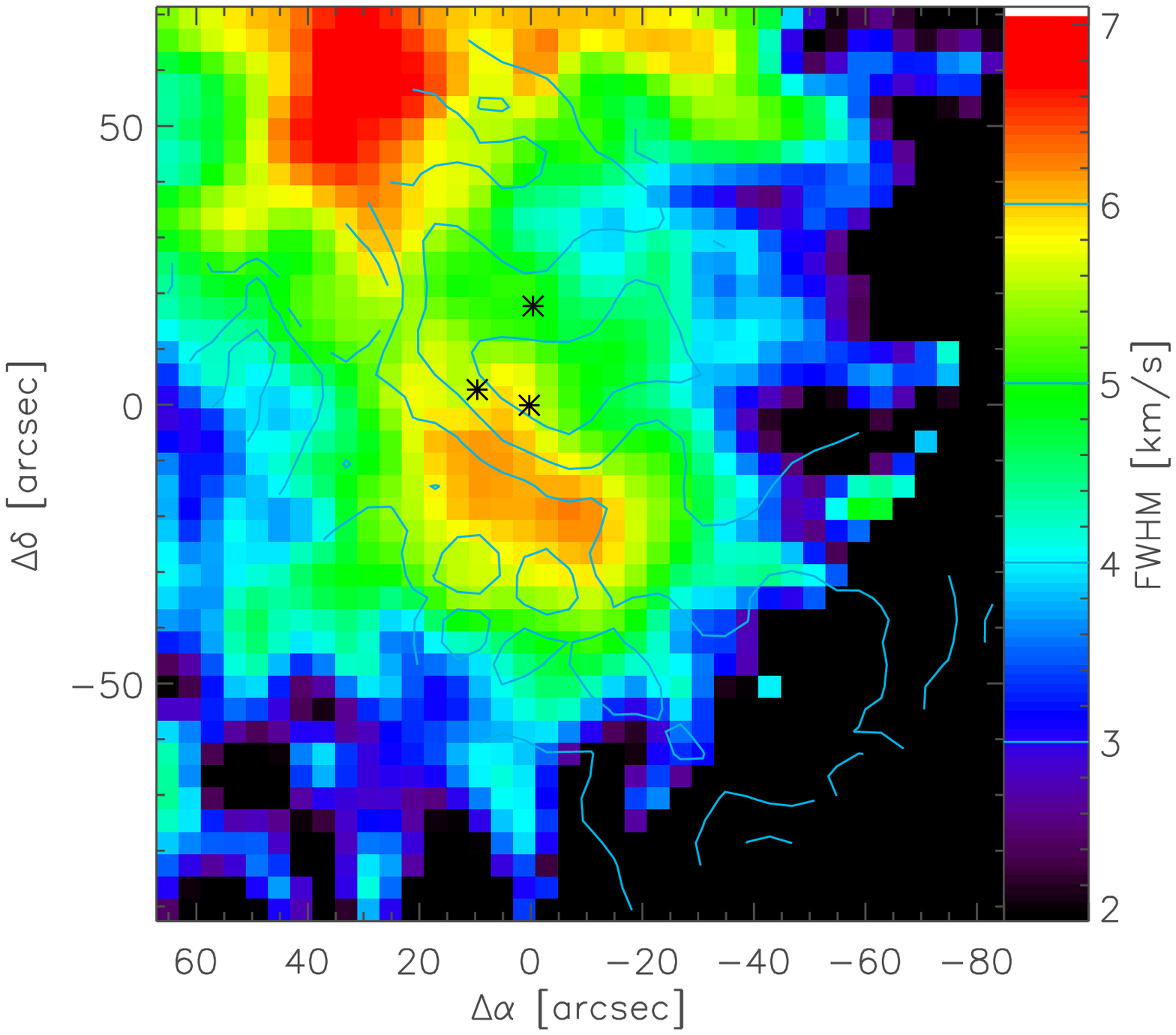}
      \caption{Peak intensity, first moment (centroid velocity) and
		second moment (translated to FWHM) of the lines of 
        $^{13}$CO 1--0 (colors) and \CII{} contours.
		The black marks indicate the positions of IRS~1-3. The
        contour levels for the \CII{} peak intensity match those from
        Fig.~\ref{fig_cii-peakintens}, for the line position they are
        given at -9.5, -8.5, -7.5, -6.5~\kms{}, 
        and for the line width at 3, 4, 5, 6~\kms{}. Velocities
        are given in LSR and intensities in main beam temperatures.}
         \label{fig_13co-cii-moments}
   \end{figure*}

In all the published maps of S~140 only the $^{13}$CO 1--0 line 
peak intensity also has a global maximum close to IRS~2\footnote{For 
simplicity we
use the location of the fine-structure line peak at about {\changed
($\Delta$R.A.,$\Delta$ Dec$) = (0'',20'')$} 
and IRS~2 synonymously in the following as the observed spatial
deviations fall within the pointing accuracy of about 2$''$.}. 
To understand 
the nature of the fine-structure line peak it is, therefore, worth to 
compare our observations with the $^{13}$CO 1--0 data more carefully,
including the velocity structure.
$^{13}$CO 1--0 is optically thin over most parts of the 
map. \citet{Koumpia2015} measured optical depths of about 0.3 for
most of the gas and a maximum value of 0.8 that is still marginally
optically thin.  Figure~\ref{fig_13co-cii-moments} 
compares the distribution of the peak intensity, the mean line
velocity, and the line width of the $^{13}$CO 1--0 line with 
the corresponding parameters from the \CII{} line. 

The relative proximity of the $^{13}$CO 1--0 peak and the \CII{} and \OI{}
peaks seem to be accidental as the overall morphology of the line maps
shows no other similarities. The  $^{13}$CO 1--0 emission is much more extended
and has about the same level at the positions of IRS~2 and IRS~1.
When inspecting the $^{13}$CO 1--0 integrated intensity, we find 
the same behavior as seen in the other molecular line tracers with a maximum 
close to IRS~1. A second similarity in the peak intensities of $^{13}$CO 1--0
and \CII{} are the relatively low values south of IRS~1. The spot matches
the southern part of the molecular outflow discussed e.g. by \citet{Maud2013}.
The outflow is driven from IRS~1 pointing towards the north-west and
south-east, leading to the velocity gradient seen e.g. in the $^{13}$CO 1--0
line position map around IRS~1. At the location of the southern outflow
\CII{} is hardly present at all, but the $^{13}$CO 1--0 line is  
prominent, tracing the outflow through broad line wings 
(see Fig.~\ref{fig_13co-cii-moments}c) that lead to a large integrated 
intensity there in spite of the lower peak intensity. 

The \CII{} line shows the largest line width north-west
of IRS~1, in the direction of the northern outflow, but only a
moderate line width at the location of the intensity peak close to
IRS~2. We find 
a trend of anti-correlation between the \CII{} lines and the $^{13}$CO 1--0 
lines. The $^{13}$CO 1--0 lines are broad in a ridge south-east and north-east
of the cluster that is associated with low center velocities while \CII{} shows
the broadest lines west of the cluster.
Except for the southern outflow region and the north-eastern
boundary of the map, the \CII{} profiles are always somewhat broader
than the $^{13}$CO 1--0 profiles. We interpret this as being due
to optical depth broadening (see below). 

When inspecting the line positions, the $^{13}$CO 1--0 profiles are
dominated by the outflow pattern with slightly blue shifted velocities
south-east of IRS~1 and red shifted velocities in the opposite 
direction\footnote{Follow-up investigations should resolve whether
the broad, blue-shifted $^{13}$CO emission north-east of
the cluster traces a second, so far unknown molecular outflow.}.
Overall, the variations are relatively small. In contrast,
\CII{} shows a clear offset in the line velocity for the
bright emission component close to IRS~2. While in most of the map
the \CII{} line has the same velocity of about $-8$~\kms{} as
the bulk of the $^{13}$CO gas, the peak source has a significantly
lower velocity of about $-6.5$~\kms{}. This indicates that the 
bright emission source is only weakly associated with the rest of the gas
and not seen in $^{13}$CO. 

   \begin{figure*}
   \centering
   \includegraphics[angle=90,width=5.95cm]{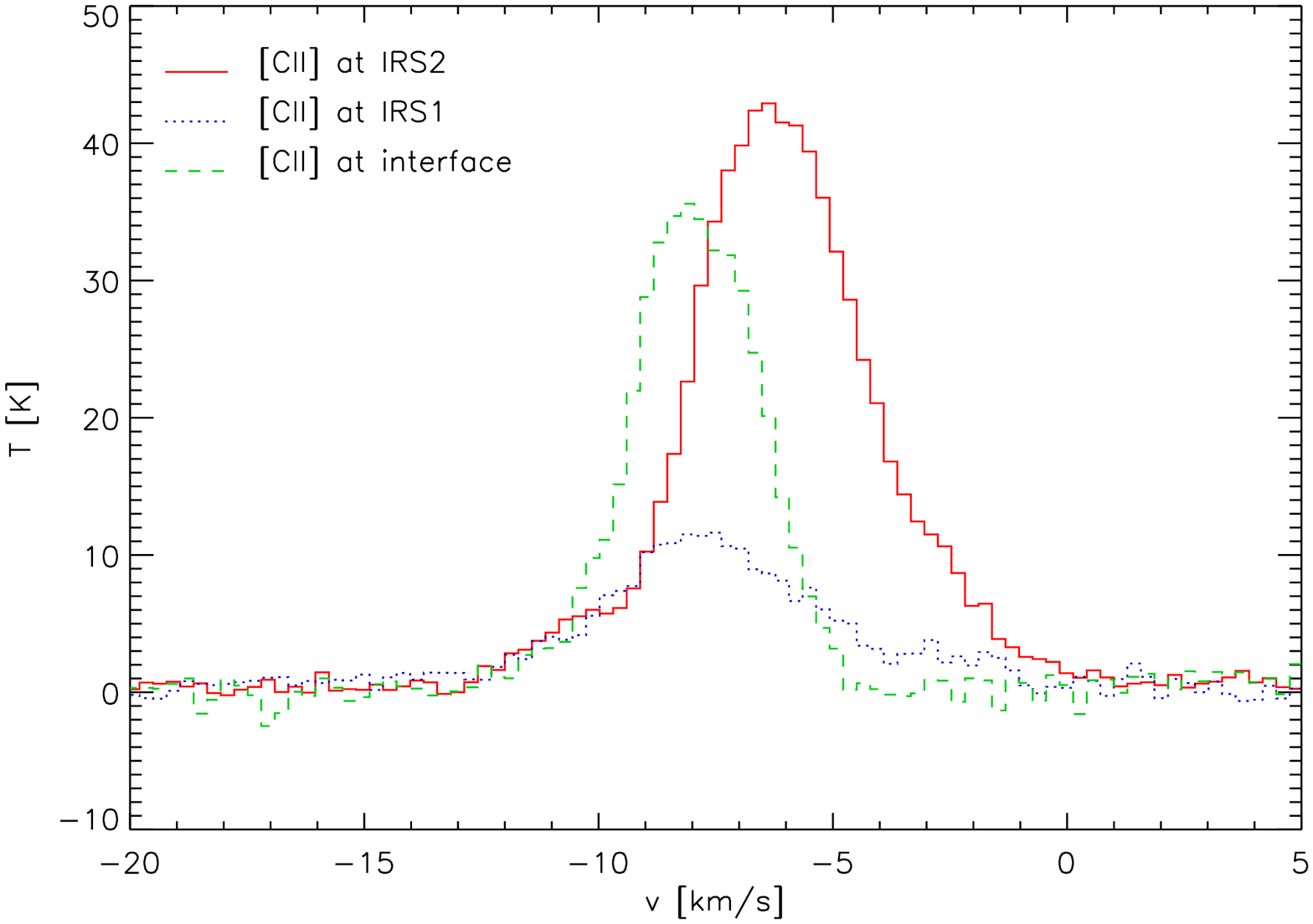}
   \includegraphics[angle=90,width=5.9cm]{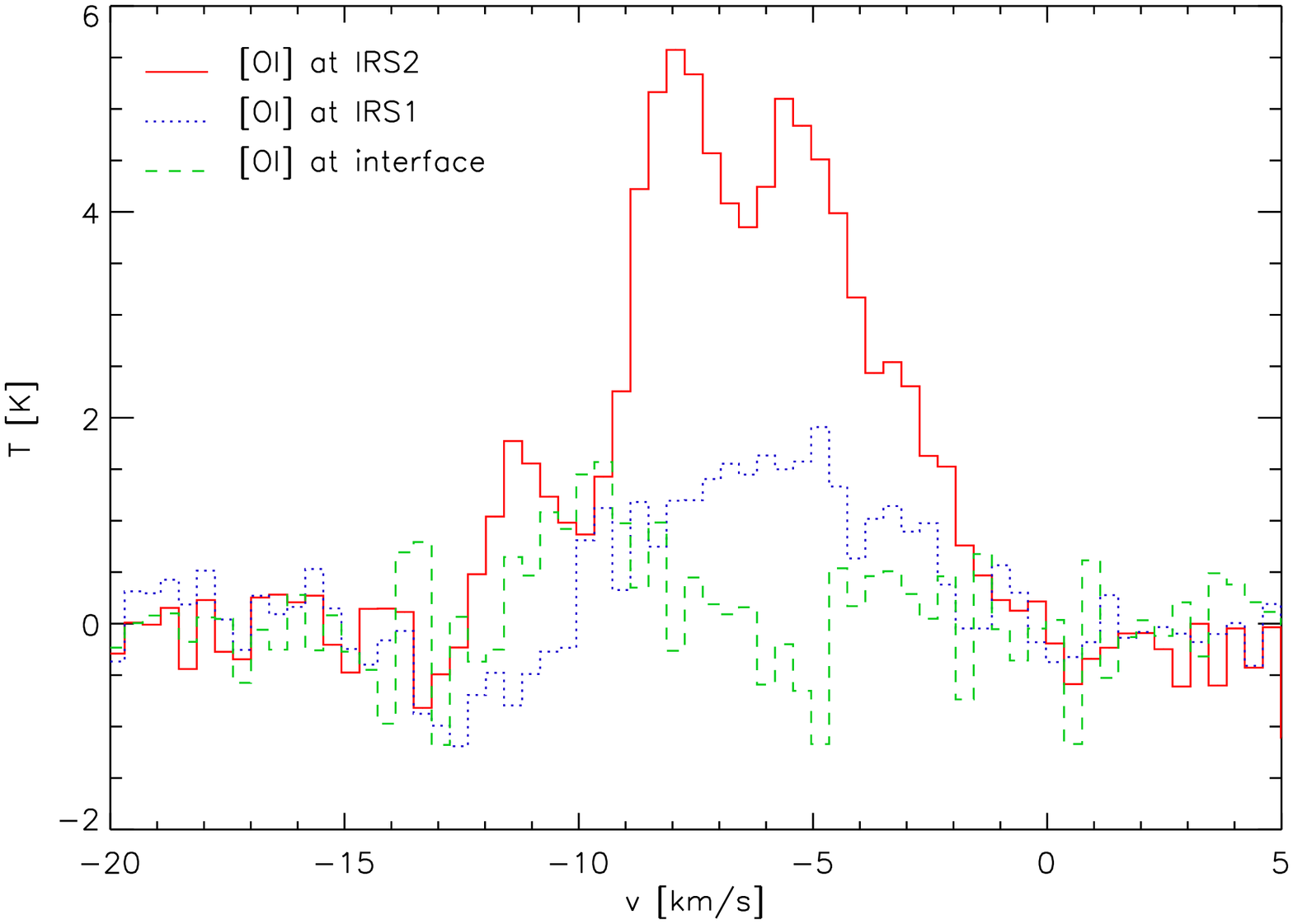}
   \includegraphics[angle=90,width=5.95cm]{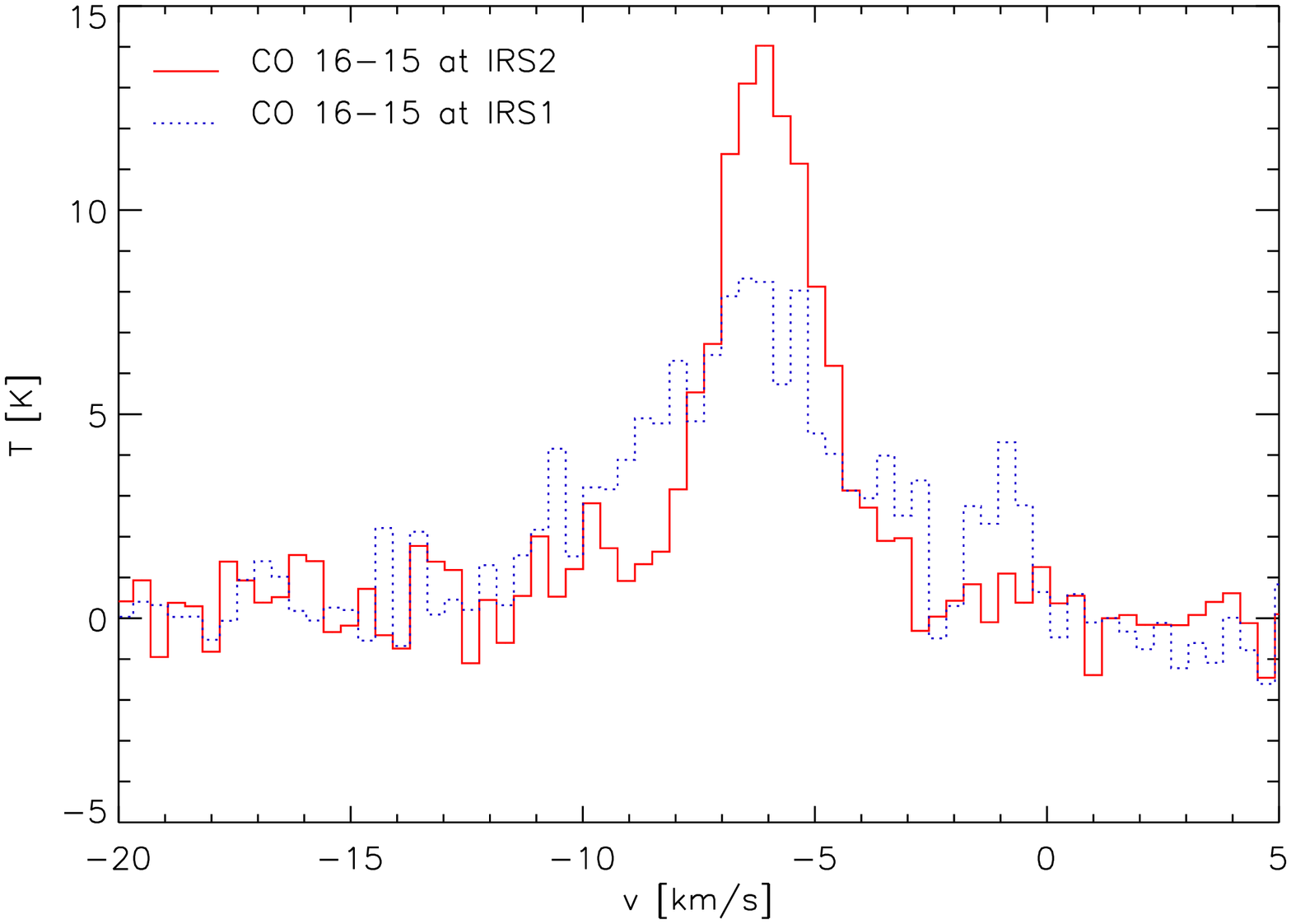}\\
   \includegraphics[angle=90,width=5.95cm]{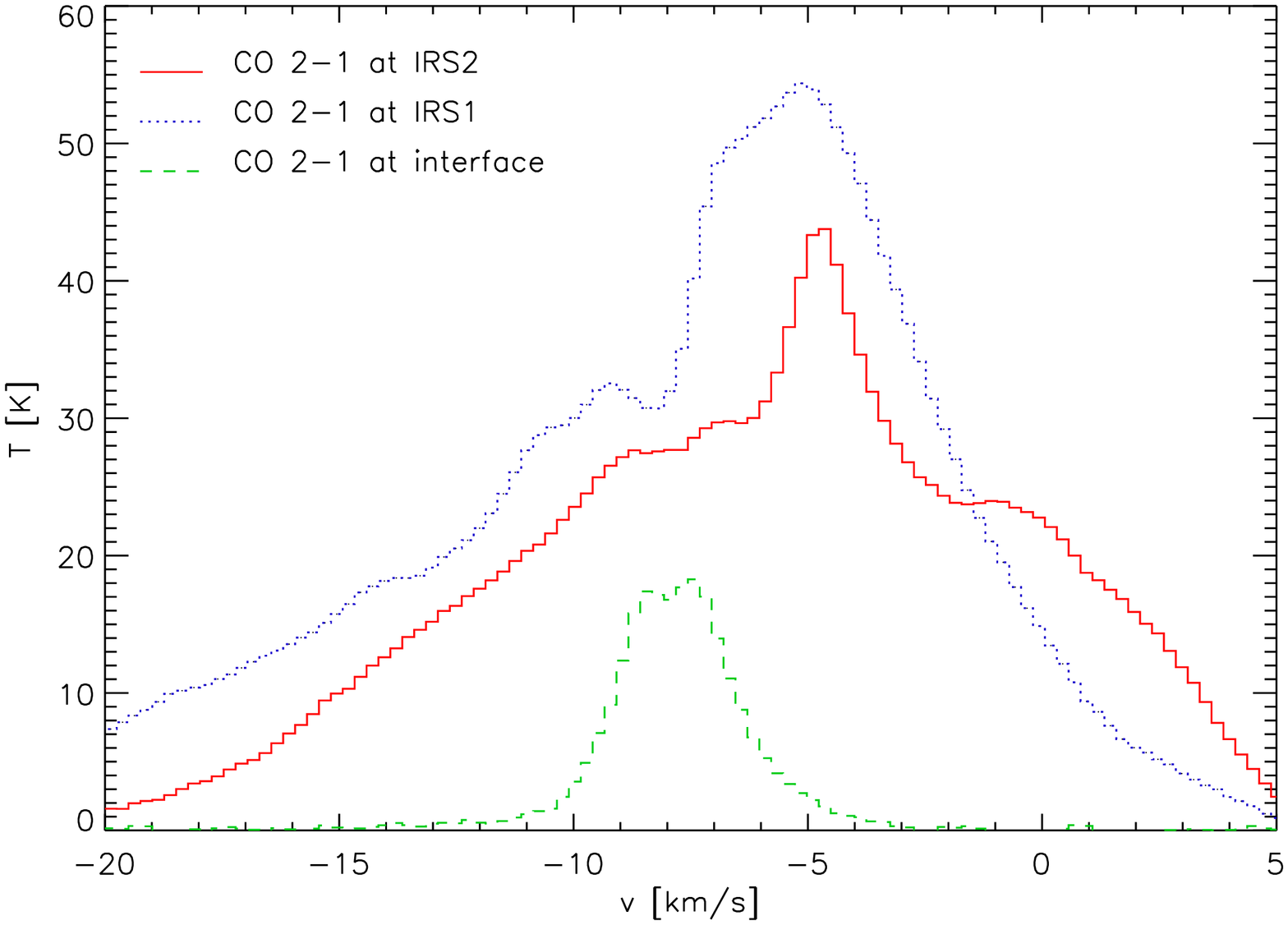}
   \includegraphics[angle=90,width=5.9cm]{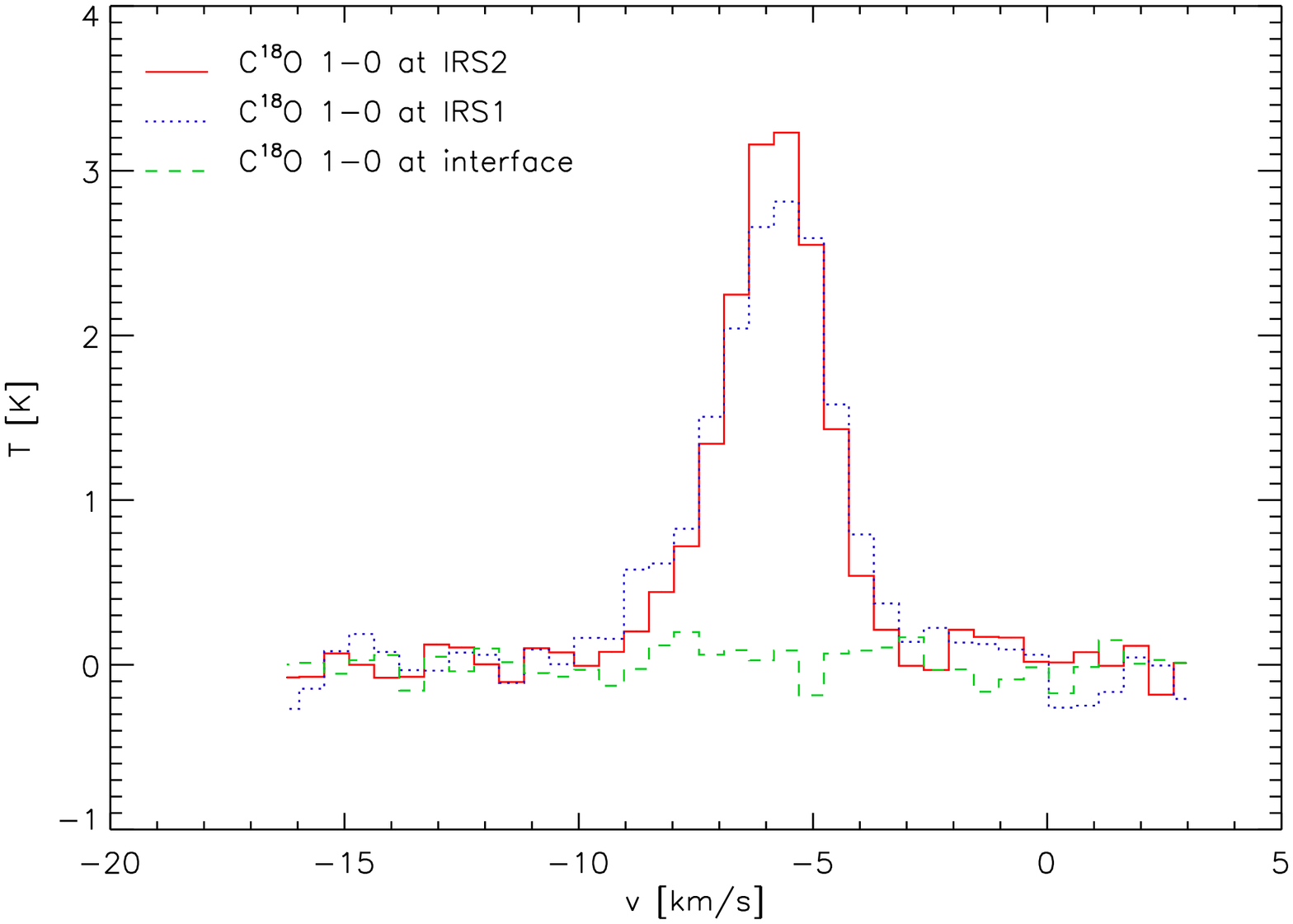}
   \includegraphics[angle=90,width=5.95cm]{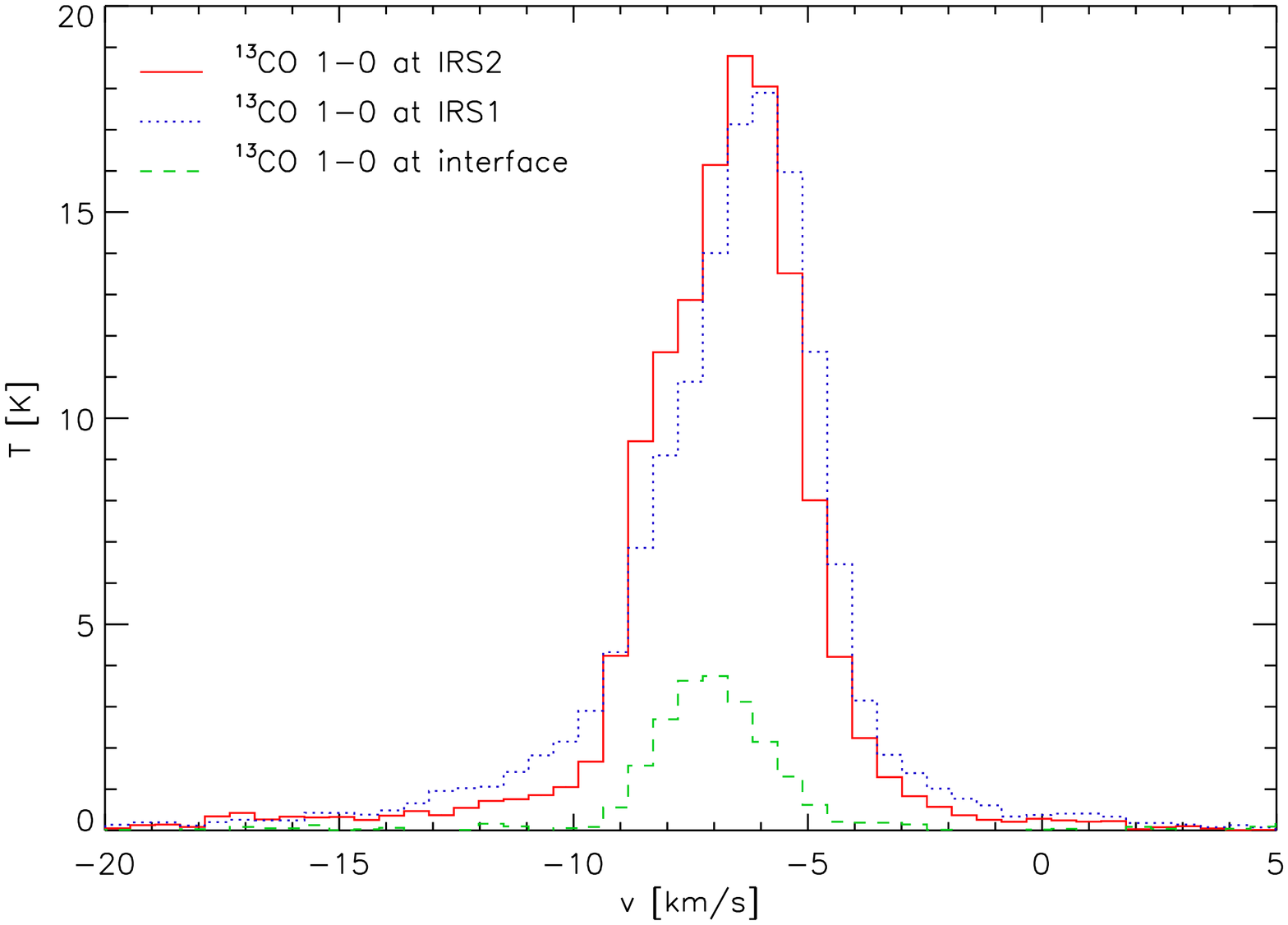}
      \caption{Line profiles of the \CII{}, \OI{}, 
        CO 16--15, CO 2--1, C$^{18}$O 1--0, and $^{13}$CO 1--0
		lines at the position of the \OI{} peak (0,20 close to IRS~2), 
        at IRS~1 (0,0), and at the interface (-40,-54).
		The lines are averaged in a Gaussian beam of 17.3$''$
		FWHM (telescope + convolution kernel) except for the 
        $^{13}$CO and C$^{18}$O lines that can be only treated at the 
        lower resolution of 23$''$.}
         \label{fig_profiles}
   \end{figure*}

Figure~\ref{fig_profiles} shows the line profiles of \CII{}, 
\OI{}, CO 16--15, CO 2--1, C$^{18}$O 1--0 and $^{13}$CO 1--0
at the location of the fine-structure line peak close to IRS~2,
at the position of the
main heating source IRS~1, and at the external cloud interface. 
All lines trace the interface at a velocity of about $-8$~\kms{}. The
\OI{} line is only marginally detected there, \CII{} and the CO 
lines are narrow with a width of 3--4~\kms{}. The velocity of this
PDR surface is slightly different from the large scale velocity field 
of the thin surrounding medium measured through the broad \HI{} line 
in the $36'$ beam of the Leiden/Dwingeloo survey to peak at 
-7~\kms{} \citep{Kalberla2005}.

Towards IRS~2, \CII{} peaks at -6.5~\kms{}, offset in velocity from 
the bulk of the gas, as seen already in Fig.~\ref{fig_13co-cii-moments}.
The \OI{} line shows a dip at this velocity and
two peaks at the velocities of
$-5.5$~\kms{} and $-8$~\kms{} coinciding with the peak velocities seen
in CO 2--1 towards IRS~2 and towards the interface. However, as
there is only weak \OI{} emission from the interface, we think
that this coincidence is accidental and we rather see in \OI{} the
same material that we trace in \CII{}, indicating that \OI{} exhibits
a self-absorption dip where \CII{} peaks instead of two velocity
components. 

CO 2--1 traces mainly other gas. We find broad outflow wings from
the molecular outflow \citep{Maud2013}, only seen in the low-$J$ CO 
lines. The optically thin lines of CO 16--15 and $^{13}$CO 1--0 are
narrow at the peak suggesting that even the \CII{} line is broadened
by a significant optical depth. The comparison of the fine structure
lines and CO 16--15 at IRS~1 shows that \CII{} only traces the
velocity component matching the interface velocity there, but no
emission at $-6$~\kms{}, while CO 16-15 and \OI{} show a stronger
emission in the $-6$~\kms{} component and a weaker emission with the
interface velocity there. This means that we find several gas
components with different velocities and chemical properties
towards IRS~1.

   \begin{figure}
   \centering
   \includegraphics[angle=90,width=\hsize]{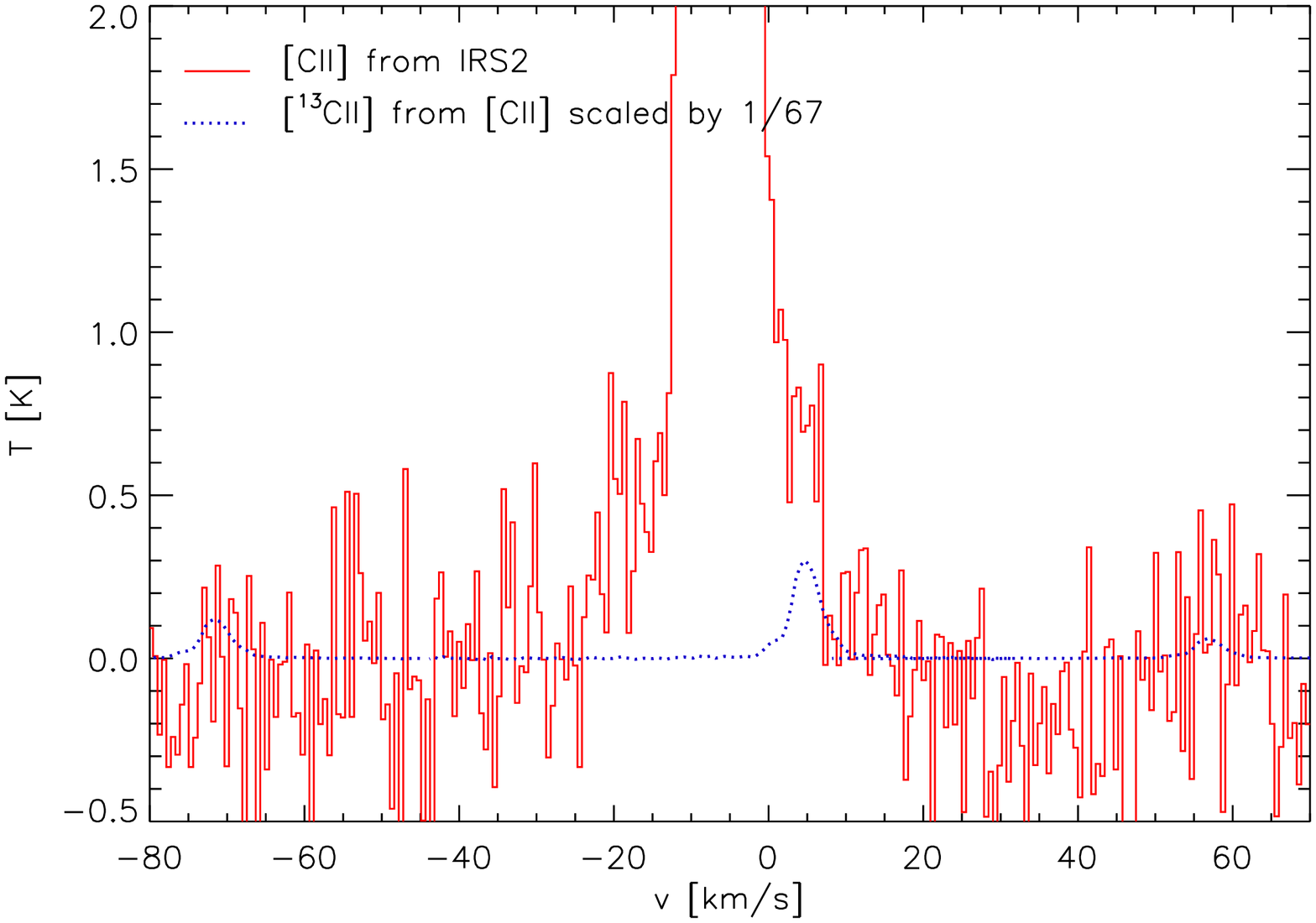}\\
   \includegraphics[angle=90,width=\hsize]{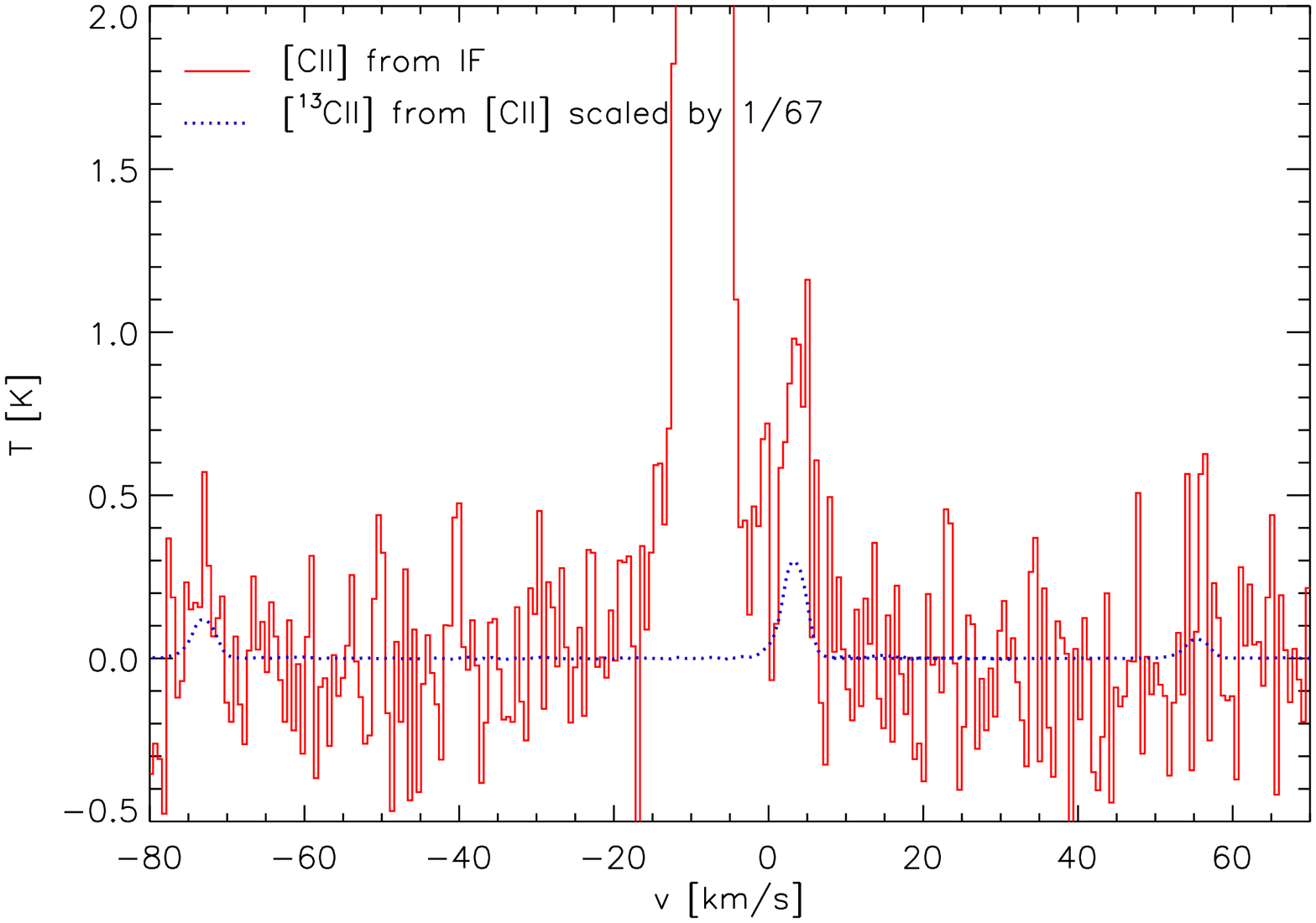} 
      \caption{Baseline of the \CII{} profiles showing the
      hyperfine components of the \thirteenCII{} transitions.
      The upper plots shows the sum spectrum from all spectra with
      integrated intensities above 100~K~\kms{} from the IRS~1-3
      region, the lower plot from the interface region. The
      blue lines represent the same spectra scaled by an
      abundance ratio of 1/67 and the corresponding line
      ratio and frequency of the three hyperfine transitions, 
      i.e. they represent the spectrum that we expect in case 
      of an optically thin \CII{} line.}
   \label{fig_13cii-spectra}
   \end{figure}

To estimate the optical depth of the \CII{} line, based
on the line profiles, we can use two approaches: derive the optical-depth
line broadening by comparing the line width with an optically thin
tracer, and to compare the line intensity with an optically thin tracer
not affected by abundance or excitation uncertainties, namely
the \thirteenCII{} transitions. In the first approach, we assume 
a Gaussian velocity dispersion and obtain the optical depth at the 
line peak from the measured line broadening $\sigma\sub{line} \approx
(1+0.115\hat{\tau})\; \sigma\sub{vel}$ where $\hat{\tau}$ is the line-center 
optical depth \citep{Phillips1979, Ossenkopf2013}. Using the
optically thin $^{13}$CO 1-0 to measure the true velocity dispersion,
we can translate Fig.~\ref{fig_13co-cii-moments}c into a \CII{} 
optical depth map, except for the region south of IRS~1 where
the $^{13}$CO 1--0 line is broader than \CII{} due to the
molecular outflow. For the interface, we obtain a $^{13}$CO 1--0 line
width of 3~\kms{} and a \CII{} line width of 4~\kms{}, in the region between
IRS~1 and IRS~2, the line widths are 5 and 6~\kms{}, respectively.
This translates into \CII{} peak optical depths of three towards
the interface and two towards the central cluster.

Figure~\ref{fig_13cii-spectra} shows the corresponding \thirteenCII{}
spectra measured in the two regions. They are compared with a
scaled version of the \CII{} line, reduced in intensity by the
typical $^{13}$C/$^{12}$C abundance ratio of 1/67 in the solar
neighborhood \citep{LangerPenzias1990,LangerPenzias1993} and the
relative weights of the hyperfine components \citep[0.625 at 11~\kms{},
0.25 at -65~\kms{}, 0.125 at 63~\kms{} relative to 
\CII{}, ][]{Ossenkopf2013}. Towards the interface the two stronger 
hyperfine components are clearly detected. The peak intensity of the
\thirteenCII{} is 3.1 times as high as expected from the 
\CII{} line when assuming optically thin \CII{} emission. This
can be translated into a line-center optical depth of about three.
Towards the embedded cluster, the strongest component appears
as a shoulder of the broader \CII{} line and the two weaker components
are barely visible. The factor 2.2 difference between observed intensity
and scaled optically thin \CII{} emission is also in agreement with
a line-center optical depth of two as measured from the line 
broadening. Both approaches, therefore provide matching results
for the \CII{} optical depth towards the two peaks of \CII{}
emission.

   \begin{figure}
   \centering
   \includegraphics[angle=90,width=\hsize]{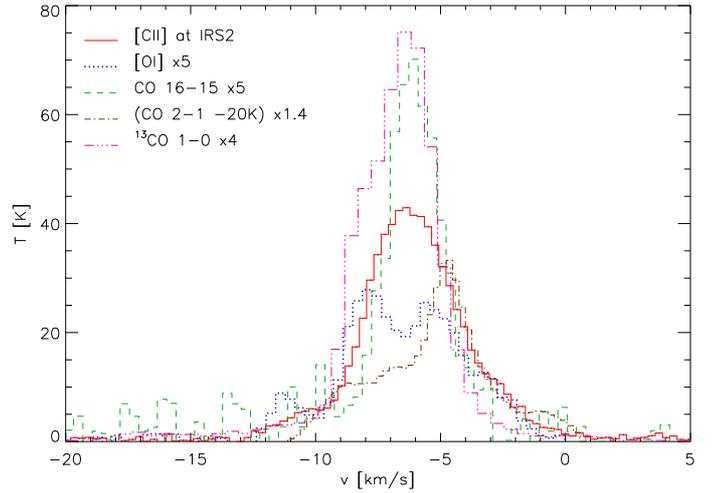}
      \caption{Line profiles of the \CII{}, \OI{}, CO 2--1, 
        CO 16--15, and $^{13}$CO 1--0 lines at the position of the 
        \OI{} peak (0,20). The lines were scaled to similar
        intensities at -4~\kms{}. For CO 2-1, a constant emission of 20~K is subtracted to exclude the extended outflow
        emission.}
         \label{fig_irs2-profiles}
   \end{figure}
   
To resolve the intrinsic velocity profile of the main emission
source at IRS~2 we compare scaled versions of the
different profiles towards this source in Fig.~\ref{fig_irs2-profiles}.
The scaling factors in the plot were adjusted to allow for a
comparison of the red wing of the profiles at sufficiently
high velocities to not be affected by self-absorption or saturation
from high optical depths. For CO 2-1 we also had to subtract the 
very extended outflow contribution (see Fig.~\ref{fig_profiles}).
The C$^{18}$O 1--0 line is not shown here as it matches almost
exactly the shape of the CO 16--15 line.
A comparison of the blue wing of the profiles provides no information
on the source at IRS~2 as this velocity interval is heavily
affected by emission and (self-) absorption from the bulk of
the gas in S~140 at -8~\kms{}.

We find an almost perfect match of the red wing shape of the
three (partially) optically thick lines of \CII{}, \OI{}, and
CO 2--1. In contrast, the thin lines of $^{13}$CO 1--0 and
CO 16--15 are clearly narrower, not allowing to fit the same
wing. Optical depth broadening of the lines must play
a significant role. The peak of the CO 16--15 line coincides
with the absorption dip seen in \OI{} and CO 2--1. The narrow 
shape of the peak of the CO 2--1 line at -4.7~\kms{} results
from this absorption out of a broader emission component.
The shoulder seen in the $^{13}$CO 1-0 and CO 2--1 lines at -8~\kms{}
probably traces emission from IRS~1 or the extended molecular
cloud material. Overall, the lines are consistent with the
velocity dispersion as measured by the CO 16--15 and C$^{18}$O 1--0
lines of only 2.4~\kms{} (FWHM), but an optical-depth broadening 
of the fine-structure lines.

For the optically thick lines, the intrinsic line intensity from the
emission peak therefore should be higher than our observed integrated
intensity. For \CII{}, the profiles are consistent with the optical
depth of about two, leading to a 45\,\% reduction of the integrated
line intensity compared to optically thin emission. For \OI{}, we
see, however, a clear self-absorption signature due to gas in the
foreground with lower excitation temperature. That makes an estimate 
of the intrinsic intensity from the emission peak difficult. If
we assume that the red wing is not affected by self absorption, we 
can use the ratio of the integrated line intensity between 
\CII{} and \OI{} seen in Fig.~\ref{fig_irs2-profiles} as an 
estimate for the missing flux in the \OI{} line. This indicates
that the source intrinsic line flux should be higher by about 50\,\%
without foreground absorption.

From the information given in the line profiles we can already conclude
that the \OI{} and \CII{} emission does not seem to be related to the
known outflow activity in S~140, probably rather tracing PDR interfaces.
While \CII{} and \OI{} are both bright at IRS~2, only \CII{} is bright 
at the cloud surface and \OI{} is barely detected there. 
As the \OI{} line has a critical density of $5\times 10^5$~cm$^{-3}$
\citep[][see Tab.\ref{tab_greatlines}]{Jaquet1992},
100 times the critical density of the \CII{} line, this probably
reflects the lower density of the interface region.
In contrast, at the fine structure
line peak position, the gas must be dense and optically thick for \OI{}
so that the same material produces an almost Gaussian line in \CII{}
and a double-peak-shaped line in \OI{}. The optically very thick
CO 2--1 line only reveals a prominent red shifted wing towards IRS~2.
This would be consistent with an expansion scenario of the source.


\section{Properties of the gas}

\subsection{Gas parameters across the GREAT map}
\label{sect_radex_for_map}

\begin{figure}
   \centering
   \includegraphics[angle=90,width=7cm]{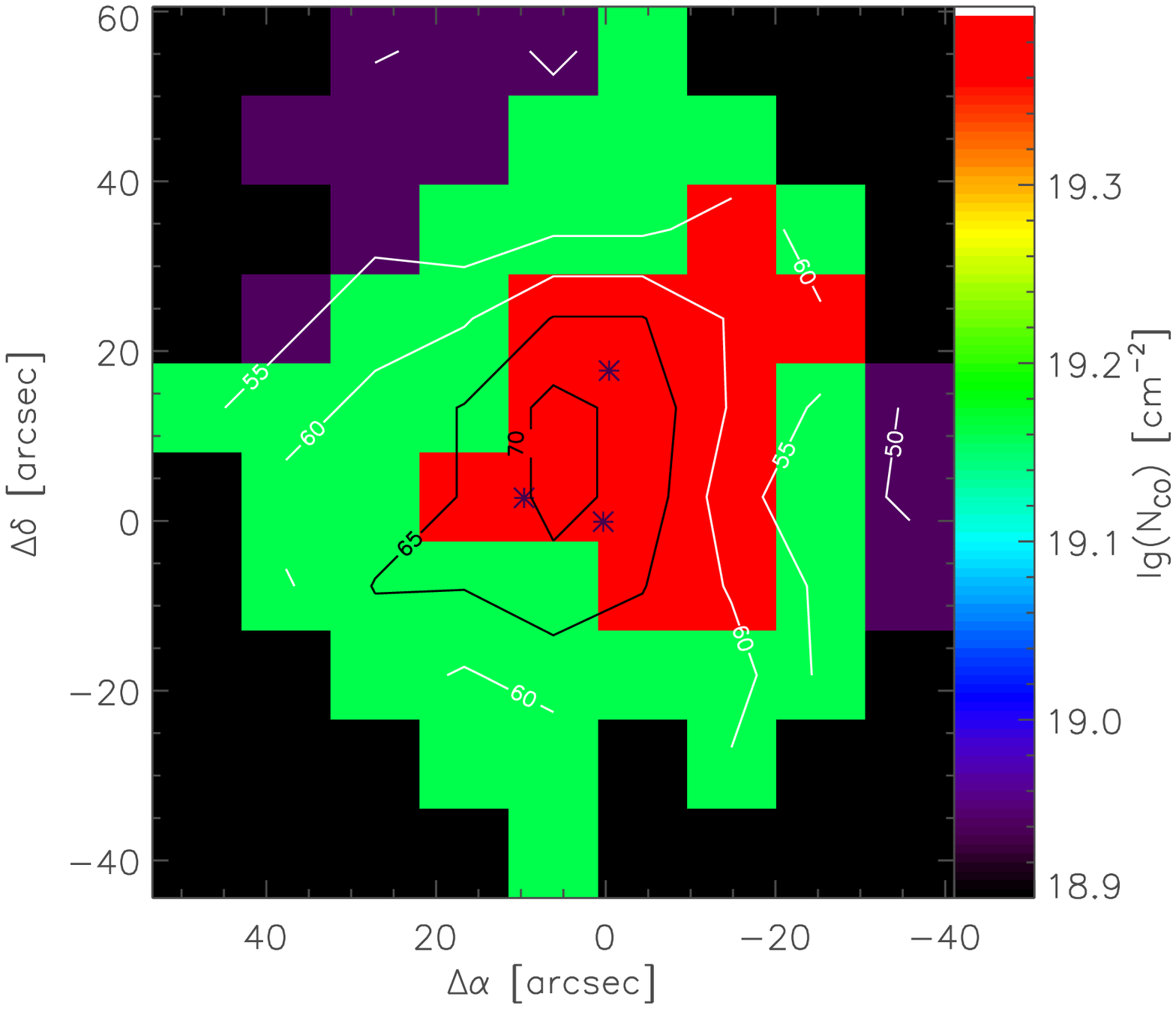}\\
   \includegraphics[angle=90,width=7cm]{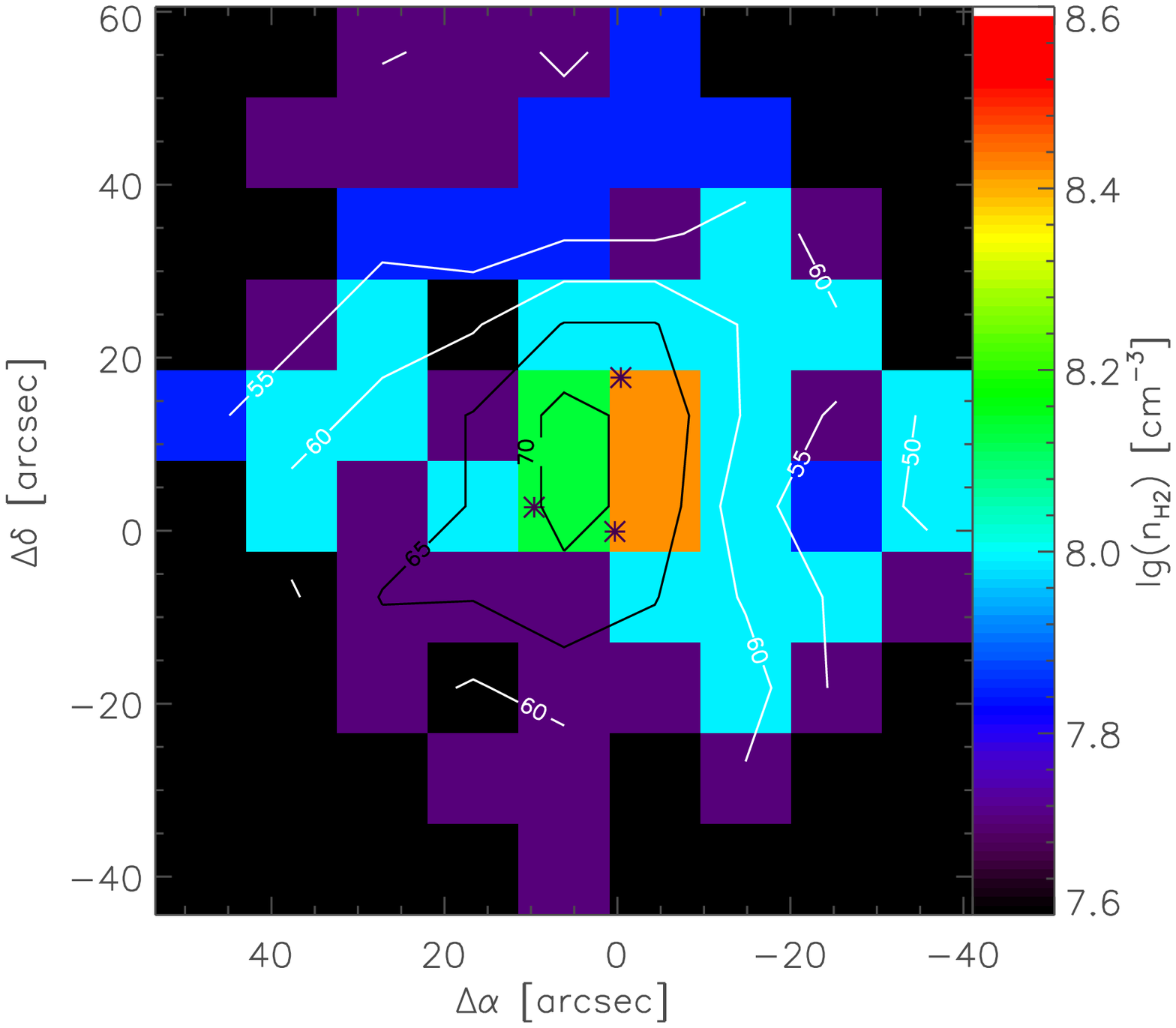}
      \caption{CO column density (upper plot: colors), gas density (lower plot: colors), and kinetic temperature (both plots: contours) derived
        from the RADEX fit of the integrated intensities of the observed
        lines of CO 1--0, 2--1, 16--15 and the isotopologues $^{13}$CO 1--0 and C$^{18}$O 1--0.}
         \label{fig_RADEX_map}
\end{figure}

As a first step to understand the nature of the fine structure lines 
we need to know the properties of the gas across the map. This can be 
done by analyzing the molecular lines of CO and its isotopologues
based on our high-$J$ CO observations and the complementary low-$J$
CO, $^{13}$CO, and C$^{18}$O IRAM data from \citet{Koumpia2015}. 
When knowing the column density, the volume density and kinetic 
temperature of the gas we can compute column density maps of 
\CII{} and \OI{} from the measured fine-structure line intensities. 

We performed a fit of the CO (and isotopologues) line intensities
using the non--LTE radiative transfer code RADEX \citep{vanderTak2007} that computes the line intensities of a molecule for a set of physical parameters: kinetic temperature, H$_{2}$ gas density, molecular column density, background temperature and line width. We used the maps of CO 1--0, 2--1, $^{13}$CO 1--0, C$^{18}$O 1--0, and CO 16--15 to perform a $\chi^{2}$ minimization fit for 
a range of kinetic temperatures (30--700~K), column densities (10$^{17}$ 
-- 4$\times$ 10$^{19}$~cm$^{-2}$) and volume H$_{2}$ densities (10$^{5}$ 
-- 5$\times$ 10$^{10}$~cm$^{-3}$). In the area covered by the
CO 16--15 map we have enough observational constraints to determine 
those 3 free fitting parameters. For all calculations  we used a fixed 
line width value of 3.5~km~s$^{-1}$. We adopted a cosmic background 
radiation field of 2.73~K and used the molecular data from the LAMDA 
database \citep{Schoier05}.

Figure~\ref{fig_RADEX_map} shows the resulting distributions of the
gas temperature, density, and column density of CO. As the CO 16--15
line has an upper level energy of 750~K and a critical density of
$10^6$~\pccm{} (see Table~\ref{tab_greatlines}), one would expect that the
strong line drives the solution to high temperatures and densities
in the order of the critical density. However, it turns out that
the low-$J$ lines can only be fitted with lower temperatures. To
excite the CO 16-15 line, RADEX then has to increase the gas density
to values in the order of $10^8$~\pccm{}, a value much higher than
derived in all other ways. This is a strong indication that the
assumption of uniform gas parameters within the individual pixels of 
our map is not justified in S~140. It is much more likely that the
true gas composition consists of a cooler component, mainly traced by 
the low-$J$ CO (and isotopologues) lines, while a second hot component
is responsible for the CO 16--15 line.

As all other lines discussed here are insensitive to gas densities
above $10^6$\pccm{}, the overestimate of the gas density obtained
in this way should be irrelevant for the further analysis. 
The RADEX fit shows that
the gas temperature peaks between IRS~1 and IRS~2, roughly matching 
the CO 16--15 intensity map. The density structure shows a density 
peak at the positions of IRS~1 and IRS~2. The column density follows 
mainly the distribution of the isotopologues which is higher towards 
the west region of the sources. It matches roughly the column density 
structure that was determined from FORCAST, PACS, and SCUBA dust 
observations in \citet{Koumpia2015}. They also found a column density
peak about 20$''$ west of IRS~1 and IRS~2 and a smaller peak somewhat
east of IRS~2.

\begin{figure}
   \centering
   \includegraphics[angle=90,width=7cm]{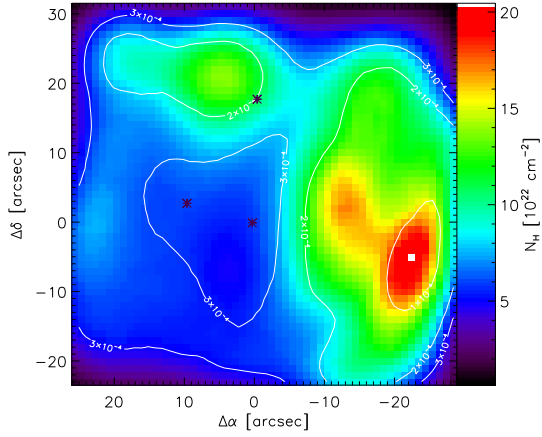}
      \caption{Column density map of gas derived from FORCAST, PACS,
	and SCUBA continuum observations
	in the center of the region (colors), overlaid by contours of
	the CO abundance, running from $1\times 10^{-4}$ to $3\times 10^{-4}$.}
         \label{fig_co-abundance}
\end{figure}

{\changed The CO column densities can be translated into abundances if
we know the gas column from the dust observations of the central region.
To derive gas column densities from the 100\micron{} dust opacity
map, we use the dust properties of model 5 from
\citet{OssenkopfHenning1994} following the discussion in \citet{Koumpia2015}
and the standard conversion factor of $N\sub{H}/A_V=1.9\times
10^{21}$~cm$^{-2}$ \citep{Bohlin1978}. This provides a 100~\micron{} optical
depth of $1.30\times 10^{-24}$~cm$^2 \times (N\sub{H}+2N\sub{H_2})$.
The resulting column density map is shown in Fig.~\ref{fig_co-abundance}.

The contours in the figure give the CO abundance computed by 
dividing the CO column from Fig.\ref{fig_RADEX_map} by the dust-based 
gas column density. For most of the cloud we find values of 
$X($CO$)=N\sub{CO}/(N\sub{H}+2N\sub{H_2})$ around $2-3\times 10^{-4}$.
Lower values of about $1\times 10^{-4}$ are found at the submm-peak
west of IRS~1. Somewhat lower values are also found towards IRS~2 and
somewhat higher values occur at the positions
of IRS~1 and IRS~3. A reduction of the CO abundance
towards IRS~2 would be consistent with the conversion of molecular material
to atomic and ionized material seen in the bright fine-structure lines.
The stronger decrease towards the submm peak could indicate CO freeze-out
in very cold and dense material. But overall, the abundance variations within 
the map are about as large
as the uncertainty from the dust column\footnote{The steep decrease
of the dust column density and the corresponding virtual increase of the CO
abundance at the boundaries of the map are probably due to missing flux
at the edges of the PACS footprint in the analysis by \citet{Koumpia2015}.}. 

{\changed A CO abundance of more than $2\times 10^{-4}$ would actually
exceed the total amount of carbon available in the gas phase \citep[see e.g.][]{Cardelli1993}.} 
\citet{Blake1987} measured values $X($CO$)$ of only $6\times 10^{-5}$ 
towards the dense molecular material in the Orion molecular cloud
and $2.5-3\times 10^{-5}$ in several clouds on larger scales, probably 
involving more atomic and ionized carbon. A higher CO abundance of 
$1.3\times 10^{-4}$ has been measured towards NGC~2024 by \citet{Lacy1994}. 
Our systematically higher abundances can be explained by
the uncertainty of the dust properties covering a factor 2--3 in the
100~\micron{} opacity \citep{OssenkopfHenning1994}. When assuming a dust
model with a two times lower 100~\micron{} opacity, the columns in 
Fig.~\ref{fig_co-abundance} increase by a factor two, the abundance 
values decrease by the same factor,
and we obtain values close to those from \citet{Lacy1994}.
Unfortunately, the resolution 
of the low-$J$ CO maps is insufficient to determine the gas column density 
with the same resolution as the dust opacity. Hence, we can only conclude that
the CO column densities are close to the upper edge of the range expected
from the dust column densities. This means that most of the gas must be
molecular so that we can use CO as a proxy for the total gas column in 
S140. This is consistent with the strong spatial confinement of the
fine-structure line emission in S140 indicating that CO provides a 
reasonable measure
for the total gas column density, even if not all gas is molecular. 
Assuming $X($CO$)=1.3\times 10^{-4}$ we obtain hydrogen column 
densities $N_{\rm H} = N_{\rm H} + 2 N_{\rm H_2}$ 
between $7\times 10^{22}$~cm$^{-2}$ and $1.97\times 10^{23}$~cm$^{-2}$
within the map covered by the CO 16-15 observations.}

\begin{figure}
   \centering
   \includegraphics[angle=90,width=7cm]{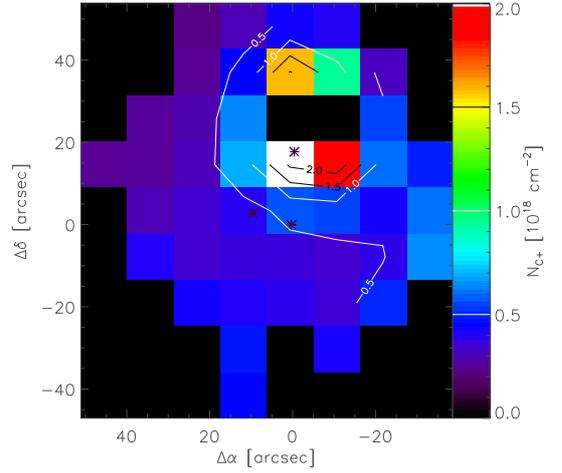}
      \caption{Column density map of C$^+$ derived from the measured
      \CII{} intensities using the gas parameters from the RADEX fit
      to the CO lines. The contours show the \CII{} line optical depth
      a levels of $\tau = 0.5, 1.0, 1.5, 2.0$.
      In the two pixels north of IRS~2, the optical depth is too large
      for a reliable column density estimate.}
         \label{fig_cii-column}
\end{figure}

\begin{figure}
   \centering
   \includegraphics[angle=90,width=7cm]{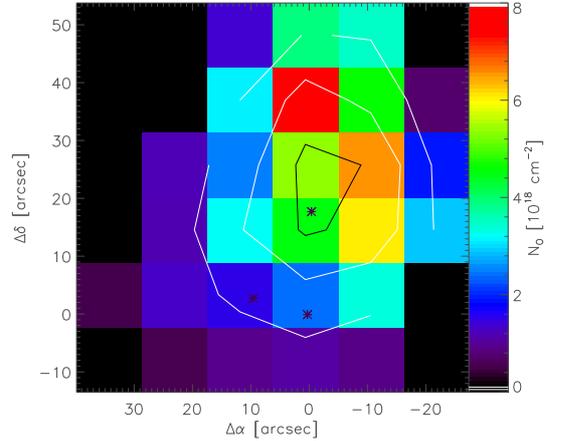}
      \caption{Column density map of atomic oxygen derived from the measured
      \OI{} intensities using the gas parameters from the RADEX fit
      to the CO lines. The contours show the \OI{} line optical depth
      at levels of $\tau = 0.05, 0.10, 0.15, 0.20$.}
         \label{fig_oi-column}
\end{figure}

{\changed In the next step we can use the kinetic temperature and volume
density from the RADEX fit of the CO lines (Fig.~\ref{fig_RADEX_map}) 
to compute the C$^+$ and atomic oxygen column densities from the
\CII{} and \OI{} observations. This allows to compare
the abundance of the different species in the gas within the telescope 
beam along the line of sight. Actually, we do not expect that all
species are spatially co-existing -- we rather expect abundance 
gradients along the path and within the area of the beam -- but the 
volume-integrated abundances can be easily compared with the results from
more complex models and other Galactic and extragalactic observations.
The CO 16--15 line provides strong constraints on the parameters
derived from the RADEX fit and it is mainly stemming from hot and dense
gas, the same gas that should also provide the main contribution to the
fine-structure line emission. Therefore, the error made in assuming the 
same excitation conditions as seen by CO for deriving the C$^+$ and oxygen
column densities should be smaller than the factor two uncertainty
that we have to face for the gas column anyway.}
To be consistent with the RADEX fit, we restrict the integration to the
same velocity interval from -8.4 to -4.4~km~s$^{-1}$.
Fig.~\ref{fig_cii-column} and Fig.~\ref{fig_oi-column} show the resulting column densities and the optical depths of the lines. Although the line profiles of \OI{} indicate optically thick line (Fig.~\ref{fig_profiles}), our resulting optical depths are $\tau < 1$. This is probably due to the strong self-absorption
in the line creating an intensity dip for the considered velocity range, 
leading to too low column densities and optical depths.
The \CII{} analysis results in column densities 2.2$\times$10$^{17}$ -- 2$\times$10$^{18}$~cm$^{-3}$ and a range of optical depths 0.28--2.31 with the higher values to be closer to IRS~2 in agreement with the estimate
from the line profile in Sect.~\ref{sect_line_profiles}.

With the hydrogen column densities from CO the C$^+$ column density map 
translates into C$^+$/H abundances, $X($C$^+)$, between {\changed $1.1\times 10^{-5}$ around
the fine-structure line peak and $2-3 \times 10^{-6}$} for the
rest of the map. For the oxygen abundance, we find 
values {\changed between $4\times 10^{-5}$ around IRS~2 and $1.0-1.5\times 10^{-5}$} 
for the rest of the map. For both species the abundance
at IRS~1 is as low as in the rest of the map outside of the 
fine-structure line emission peak.

These values can be compared to PDR models \citep{Roellig2007}. 
At a PDR surface, almost all carbon is in the form of C$^+$. For 
diffuse clouds, \citet{Sofia2004} measured a C$^+$ abundance of 
$X($C$^+)=1.6\times 10^{-4}$, but in dense regions 
a larger fraction of carbon is incorporated in dust and 
PAHs leading to a {\changed PDR surface abundance of $X($C$^+)=1.2\times 10^{-4}$
in dense clouds} \citep[][]{Wakelam2008}. Deeper in the clouds where the UV is 
sufficiently shielded, C$^+$ turns into atomic carbon and 
subsequently into CO, so that the C$^+$ abundance falls by 
three to four orders of magnitude, depending on the cosmic ray 
ionization rate. Even our value for IRS~2 falls much below the
PDR {\changed surface limit, confirming that only in a small gas fraction 
carbon is ionized through UV radiation while most
of the gas is molecular. The} column of molecular material should be 
{\changed at least ten} times
deeper than the C$^+$ column produced in a PDR around IRS~2.

Depending on the local conditions the oxygen abundance in dense clouds
should not vary by more than a factor three. In the 
high-UV field limit, almost all gas-phase oxygen is in atomic oxygen, 
resulting in an abundance of $X($O$) = 3\times 10^{-4}$ relative to 
hydrogen. Deep in molecular clouds, most oxygen is incorporated in CO 
so that only $X($O$) = 1\times 10^{-4}$ is left in the gas phase. 
Our measured abundance still falls below this lower limit. This can be
explained by the fact that most oxygen is probably at temperatures 
below the range of 50--70~K fitted by RADEX for a homogeneous gas.
The colder oxygen does not contribute to the fitted emission, but rather 
creates the absorption seen in our \OI{} line profiles.

\subsection{The \CII{} and \OI{} peak}
\label{sect_irs2}

\begin{figure}
   \centering
   \includegraphics[angle=90,width=6.5cm]{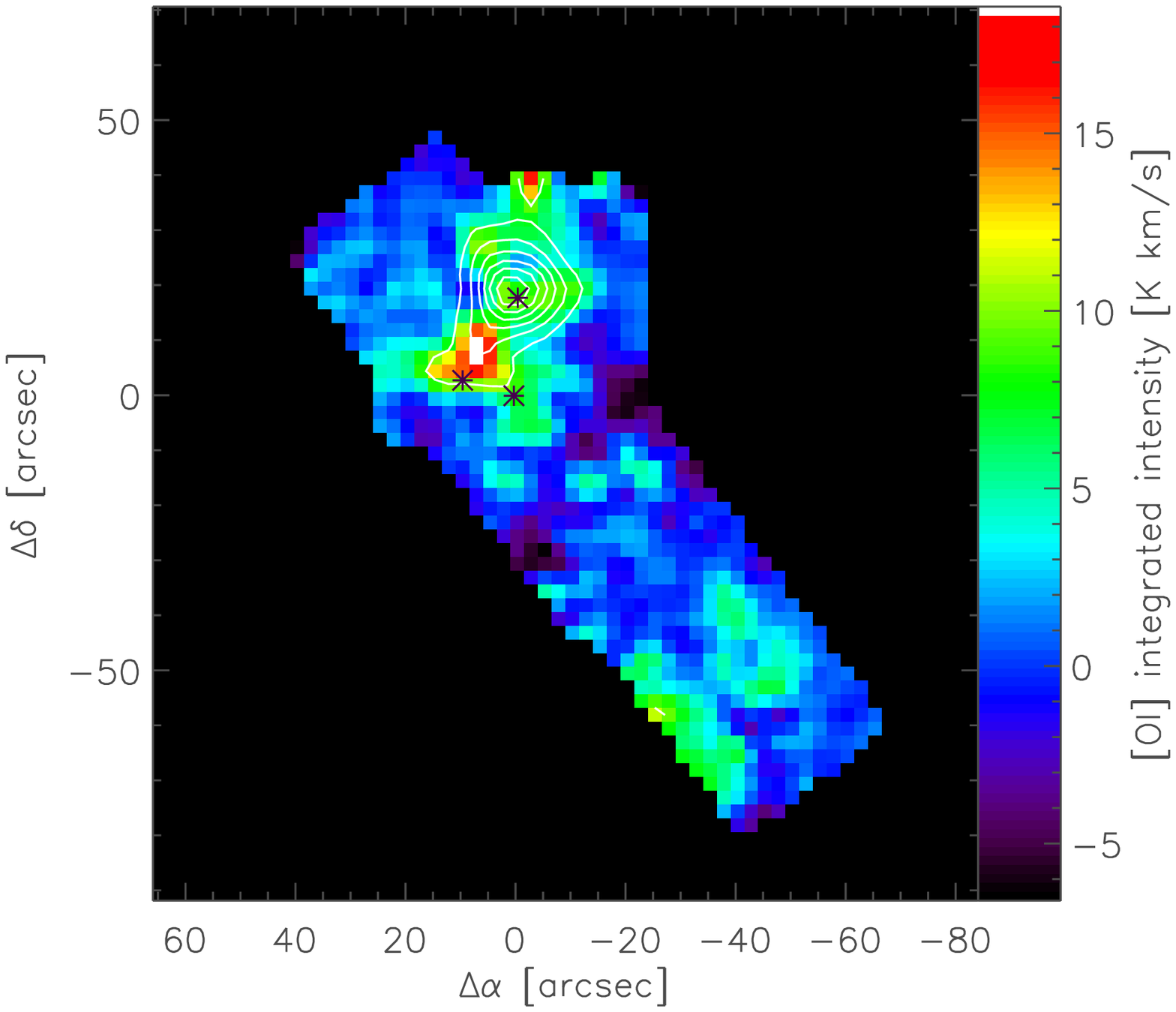}\\
   \includegraphics[angle=90,width=6.5cm]{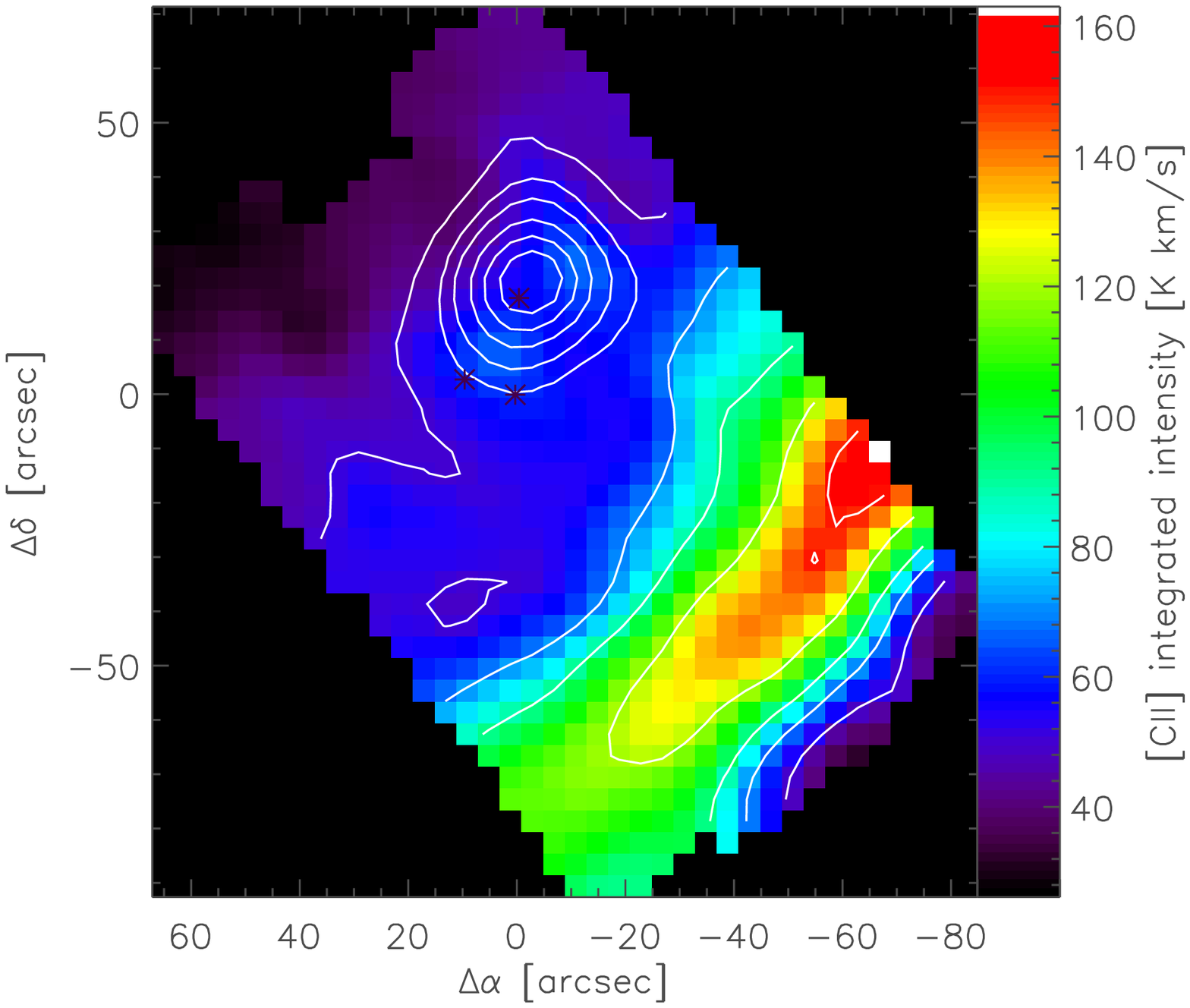}\\
   \includegraphics[angle=90,width=6.5cm]{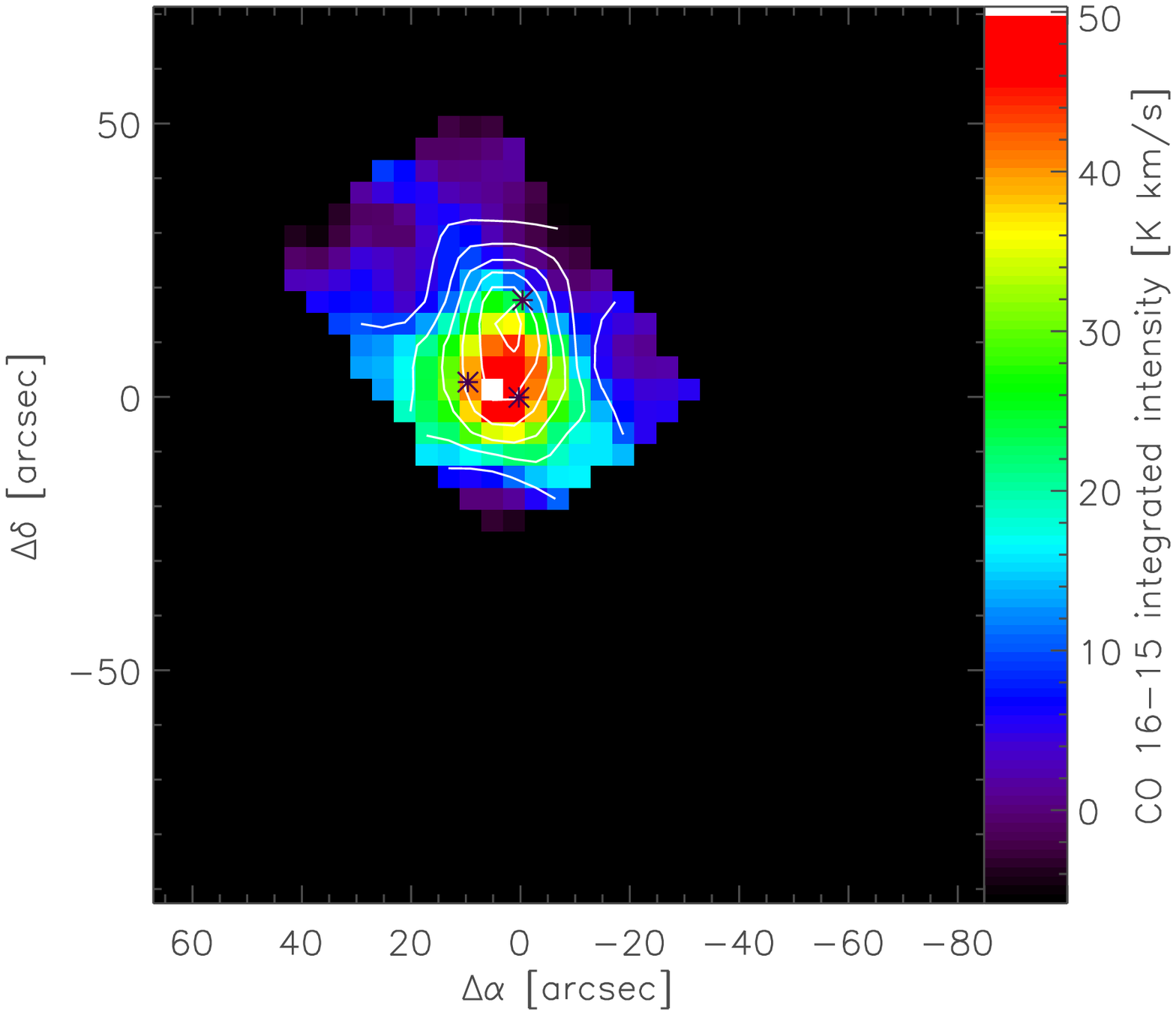}
      \caption{Intensity map of \OI{} (top), \CII{} (center),
        and CO 16--15 (bottom) after
		subtracting a Gaussian intensity distribution of
		FWHM$=8.3''$ from the observed map. The overlaid
		contours show the original intensity maps at levels
		of 8, 24, 40, 56, 72~K~\kms{} (\OI{}),
		50,70,90,120,150,180~K~\kms{} (CII{}), and 10, 20, 30,
        40, 50, 60~K (CO 16--15). }
         \label{fig_subtracted}
\end{figure}

To estimate the size of the region responsible for the fine-structure
line emission peak,
we fitted the observed \OI{} integrated intensity map by a Gaussian 
intensity distribution peaking close to IRS~2 convolved with
the telescope beam at 4.7~THz. The \OI{} map provides the main constraint to
the source size due to the smaller beam compared to \CII{} and the
CO lines. We allowed for a 2~arcsec pointing jitter in the fit
when adjusting the location of the peak. The best fit leading to a
relatively smooth background after the source subtraction is
obtained for a source FWHM of 8.3$''$ and a peak intensity
of 76~K~\kms{}. Figure~\ref{fig_subtracted} (top panel) shows the 
result. We see that the subtracted source covered 80\,\% of the
intensity at the position of the peak, but that there remains
a somewhat more extended emission structure north of IRS~1.
IRS~1 is part of that structure, but shows no concentrated
\OI{} emission associated with the infrared source. In contrast,
we find a secondary \OI{} peak slightly north of IRS~3, the weakest
infrared source in the field.

A priori, the spatial extent of the emission in \OI{} and \CII{}
can be different, but if we subtract a source with the 
same size and the peak intensity of 212~K~\kms{} from the \CII{} map 
we also obtain a very smooth map with no further indications of 
the IRS~2 contribution (Figure~\ref{fig_subtracted}, central panel). 
This suggests that we see about the same
gas responsible for the peak in \OI{} and \CII{} spatially well
confined to a region close to IRS~2.
Repeating the same experiment for the CO 16--15 map, provides
a peak intensity of 46~K~\kms{}, but a clear residual of 
emission from
IRS~1 and IRS~3 (Figure~\ref{fig_subtracted}, bottom panel).
This agrees with the previous analysis of \citet{Koumpia2015}
finding extended hot molecular gas around IRS~1.

The identification of a spatially well confined emission structure
close to IRS~2, responsible for the observed strong fine-structure 
line emission allows to quantify the properties of the emitting
gas. With an FWHM of 8.3$''$ we obtain a spatial angle of the
emitting region of $1.8\times 10^{-9}$~sr and a diameter of 0.03~pc.
This allows to compute absolute luminosities from the integrated
intensities and the source projected area of $1.0\times 10^{34}$~cm$^{2}$
assuming isotropic radiation. For the \OI{} line we obtain
0.28~$L\sub{\odot}$ and for the \CII{} line 0.05~$L\sub{\odot}$. 
This can be compared to the total energy input at that location.
\citet{Koumpia2015} estimated a total luminosity of the
source at IRS~2 of 2000~$L\sub{\odot}$. This means that only
0.016\,\% of the source luminosity is radiated away in the
two fine-structure cooling lines. We will further discuss this 
in Sect.~\ref{sect_cooling_balance}. 

For the CO 16--15 line we
obtain a corresponding luminosity of only 0.01~$L\sub{\odot}$,
but over the full CO ladder this will add up to a considerable
net cooling. If we make the extremely conservative assumption that
all CO line intensities are constant at the value of the 16--15
transition for lower $J$ transitions and zero for higher 
transitions, we obtain a lower limit for the CO cooling luminosity
of 0.045~$L\sub{\odot}$ already matching the \CII{} luminosity. 
If we use the full SED ladder fit to the narrowest line component
from the next section (\ref{sect_RADEX_for_sources}), we obtain 
instead a cooling power of 0.096~$L\sub{\odot}$ and if we include
the emission over the total line width, the CO luminosity sums
up to 0.13~$L\sub{\odot}$, i.e. a value that is about half the
cooling power in the \OI{} line. In spite of the
prominent fine-structure line peak at IRS~2, we thus find
a situation where the molecular line cooling is relatively
important next to the fine structure line cooling.  This
is similar to the situation found in other protostars 
\citep{Karska2013}. Obviously, the
situation is even more extreme for IRS1, where we measure
strong high-$J$ CO emission, but basically no \CII{}.

We can estimate the gas column density from the integrated
\CII{} line intensity, by using the relatively constant \CII{}
emissivity per column density \citep[Eq.~(2) from][]{Ossenkopf2013}.
For a temperature of about 200~K we obtain:
\begin{equation}
\int\!\! \epsilon\, dv \approx 560\;\frac{\rm K\,km s^{-1}}{\rm cm^{-3}~pc} 
        \times N_{{\rm C}^+}
\label{eq_emissivity}
\end{equation}
As the emissivity is only weakly temperature-dependent, this value
applies within a 15\,\% error bar to all excitation temperatures between
100~K and 500~K, covering the typical range expected in PDRs. 
Eq.~(\ref{eq_emissivity}) provides a direct translation of intensity
to column density for optically thin emission. Because we know the optical
depth from Sect.~\ref{sect_line_profiles} the equation can also be used 
with  the corresponding optical depth correction of 45\,\% to compute the
C$^+$ column density for IRS~2.

We obtain a C$^+$ column density of $1.7\times 10^{18}$~cm$^{-2}$.
This is slightly lower than the value of $2\times 10^{18}$~cm$^{-2}$ 
obtained for the total \CII{} emission towards IRS~2 measured in 
the RADEX simulation for the full map in Sect.~\ref{sect_radex_for_map}.
However, here we consider only the 73\,\% of the \CII{} emission
at that position that are attributed to the compact source ignoring
a small part of the total column density.
Using the carbon gas-phase abundance of $1.2\times 10^{-4}$ 
for dense clouds \citep{Wakelam2008} (see Sect.\ref{sect_radex_for_map}) 
and assuming that all carbon is 
ionized, the C$^+$ column density of $1.7\times 10^{18}$~cm$^{-2}$
provides a lower limit for the total column density of the
\CII{} emitting gas of $1.4\times 10^{22}$~cm$^{-2}$. Higher values
result for partial ionization.
If we assume that the \CII{}-emitting gas fills the observed
emitting volume of $1.0\times 10^{51}$~\ccm{}, we deduce a minimum gas
density of $1.4 \times 10^{5}$~\pccm{}. The derived optical depth of 
the clump seen
in dust emission close to IRS~2 is, however, $A\sub{V} \approx 100$
\citep{Koumpia2015} corresponding to gas column densities of 
$2\times 10^{23}$~\pscm{}. That means, that only 7\,\% of the gas
is emitting in \CII{}. This could be explained by clumpiness of the
gas within the 8.3$''$ peak and high-density clumps, or more naturally, 
by a PDR layering structure where the \CII{} emission only stems 
from gas with $A\sub{V} \la 2$ facing the embedded sources. 

Unfortunately, it is not possible to perform the same estimate for
the \OI{} emission. The high-temperature LTE limit of the \OI{}
emissivity is about three times as large as for \CII{}
(Eq.~\ref{eq_emissivity}), but it only applies to densities
above $10^6$~\pccm{} and temperatures above 200~K, and it also
requires optically thin emission or an accurate estimate. The 
line profiles showed already that \OI{} is optically very
thick and the intensity of the \OI{} line is three times lower 
than that of \CII{}. To check the consistency of the parameters
in the frame of a real radiative transfer computation we perform
RADEX runs in the next section.

\subsection{RADEX fit of individual sources}
\label{sect_RADEX_for_sources}

\begin{table*}[t]
\caption{Line parameters of the narrow component of CO and isotopologues towards IRS~1 and IRS~2 -- GREAT, IRAM \& HIFI data.}
\centering
\small\addtolength{\tabcolsep}{-2.0pt}
\begin{tabular}{l| c c c| c c c |c c c}
\hline
 & & \bf IRS~1 & & & \bf IRS~2 & & & interface \\
\hline\hline
Line 
& $v_{\rm LSR}$ & FWHM & $\int$ T$_{mb}dv $ 
& $v_{\rm LSR}$ & FWHM & $\int$ T$_{mb}dv$ 
& $v_{\rm LSR}$ & FWHM & $\int$ T$_{mb}dv$ \\
& [km~s$^{-1}$] & [km~s$^{-1}$] & [K~\kms{}] 
& [km~s$^{-1}$] & [km~s$^{-1}$] & [K~\kms{}] 
& [km~s$^{-1}$] & [km~s$^{-1}$] & [K~\kms{}]\\
\hline\hline
CO 9--8 
& $-5.78\pm0.01$ & $4.87\pm0.06$ & $117\pm2$ 
& \ldots & \ldots & \ldots 
& $-7.80\pm0.07$ & $2.50\pm0.20$ & $1.20\pm0.20$ \\
CO 13--12 
& \ldots & \ldots & \ldots 
& $-6.76\pm0.03$ & $3.30\pm0.08$ & $71.1\pm1.4$
& \ldots & \ldots & \ldots \\
CO 16--15 
& $-6.55\pm0.06$ & $3.5\pm0.2$ & $40.3\pm1.7$ 
& $-6.43\pm0.04$ & $2.7\pm0.1$ & $37.2\pm1.3$
& \ldots & \ldots & \ldots \\
$^{13}$CO 1--0
& $-6.45\pm0.03$ & $2.48\pm0.09$ & $28.7\pm1.9$ 
&$-6.54\pm0.06$ & $2.6\pm0.1$ & $49.4\pm1.8$ 
& $-8.12\pm0.37$ & $1.34\pm0.70$ & $0.84\pm0.26$ \\
$^{13}$CO 10--9 
& $-6.71\pm0.02$ & $2.41\pm0.08$ & $12.8\pm1.1$ 
& \ldots & \ldots & \ldots 
& \ldots & \ldots & $<3 \sigma$ \\
C$^{18}$O 1--0 
&  $-6.52\pm0.05$ & $2.4\pm0.1$ & $6.8\pm0.6$ & 
$-6.64\pm0.04$ & $2.36\pm0.1$ & $8.4\pm0.1$ 
& \ldots & \ldots & $<3 \sigma$ \\
C$^{18}$O 9--8 
& $-6.7\pm0.1$ & $2.6\pm0.2$ & $2.8\pm0.3$ 
& \ldots & \ldots & \ldots 
& \ldots & \ldots & $<3 \sigma$ \\
\CII{} $^3P_{3/2}-^3P_{1/2}$ 
& $-7.44\pm0.04$ & $5.94\pm0.09$ & $72\pm1$
& $-6.27\pm0.01$ & $4.62\pm0.03$ & $159.5\pm0.9$ 
& $-7.92\pm0.01$ & $3.24\pm0.02$ & $119.6\pm0.7$ \\
\OI{} $^3P_1-^3P_2$ 
& $-5.9\pm0.2$ & $5.3\pm0.4$ & $9.5\pm0.7$
& $-6.4\pm0.1$ & $5.9\pm0.2$ & $25.4\pm0.7$
& $-9.1\pm0.6$ & $2.7\pm0.5$ & $3.1\pm0.9$ \\
\hline\hline
\end{tabular}
 \tablefoot{The line parameters are measured
 in a 21$''$ Gaussian aperture around the indicated 
 position (telescope beam + convolution kernel) to match 
 the coarse resolution of the $^{13}$CO 1--0 map.}
\label{tab_co_lines}
\end{table*}

%
%

A more accurate estimate can be obtained when using the known
gas parameters from a fit to the observed CO transitions and taking 
the line optical depth into account in the frame of a RADEX computation.
As all lines are bright towards the central sources, IRS~1 (0,0) and 
IRS~2 (0,20), we can  avoid the low--J CO lines there that show 
complex line profiles and are very optically thick and fit only the 
higher transitions of CO and all available transitions of the 
isotopologues in RADEX.
For the interface (-40,54), the RADEX analysis was already performed in
\citet{Koumpia2015} using the low-$J$ CO lines and 
an additional HIFI cut measuring CO 9-8 and $^{13}$CO 10-9.
As our CO 16--15 observations do not cover this region,
no further constraints are provided to the fit so that we
can reuse the results obtained there.
The HIFI cut also covered IRS~1 so that we have the most complete
data set for this source.
Table~\ref{tab_co_lines} contains all transitions and line properties  
that were used for the RADEX fit and the further analysis.

\begin{table}
\caption{Physical conditions as derived with RADEX for selected positions}  
\label{table_RADEX}      
\centering                         
\begin{tabular}{l c c c c}        
\hline\hline                 
Position & $T\sub{kin}$  & $N($CO) & $N\sub{H}$ & $n\sub{H_2}$ \\
& [K] &  [$10^{18}$~cm$^{-2}$] & [$10^{22}$~cm$^{-2}$] &  [cm$^{-3}$]\\
\hline  
IRS~1 					& 75  & 6.6 & 11 & 3$\times$10$^{9}$ \\
IRS~2 					& 73 &  16 & 27 & 7.1$\times$10$^{8}$ \\
Interface    			& 40 & 1.0 & 1.5 & 4$\times$10$^{5}$ \\      
\hline                                   
\end{tabular}
\tablefoot{The synthetic integrated intensities are calculated using a FWHM of 3.2~km~s$^{-1}$ for IRS~1 \& 2.7~km~s$^{-1}$ for IRS~2. The parameters towards the IF were derived in addition the CO 1--0 \& CO 2--1 .}
\label{table_parameters}
\end{table}

\begin{figure}
   \centering
   \includegraphics[angle=90,width=\hsize]{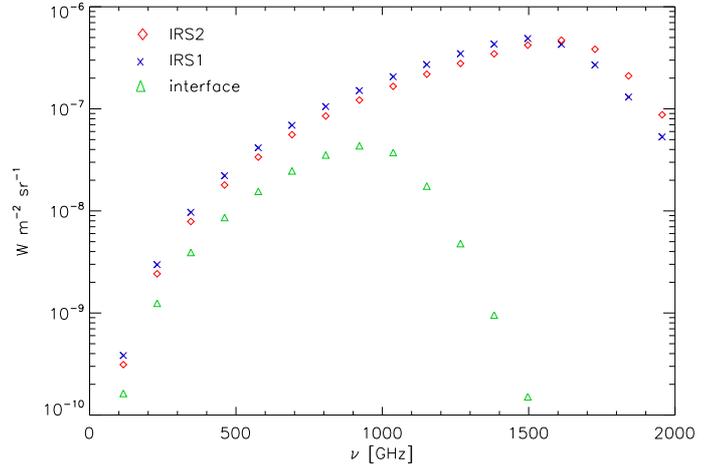}
      \caption{CO ladder from the RADEX fit to the line intensities
      at the three prominent positions.}
      \label{fig_sedplot}
\end{figure}

Figure~\ref{fig_sedplot} shows the resulting CO SED from the RADEX fit
for all lines up to $J=17-16$ and Table~\ref{table_parameters} summarizes 
the resulting gas parameters. 
In the fourth column, we translated the CO column density into an 
equivalent hydrogen column density using the abundance factor of 
$6\times 10^{-5}$ from Sect.\ref{sect_radex_for_map}. 
The higher column density towards IRS~2 compared to that of IRS~1 
is driven by the spatial distribution of the low--J CO isotopologues
and the high--J CO that is stronger towards IRS~2 (see Fig.~\ref{fig_profiles}).
The hydrogen column densities towards the two sources approximately 
match the visual extinction of $A\sub{V} \approx 50$ towards
IRS~1, and $A\sub{V} \approx 100$ towards IRS~2 measured by \citet{Koumpia2015}
when assuming a standard conversion factor of $N\sub{H}/A_V=1.9\times
10^{21}$~cm$^{-2}$ \citep{Bohlin1978}.

The IRS~1 and IRS~2 are warmer than the interface and have a similar temperature. The absolute value of the temperature is, however, questionable as it
depends on the assumption of the uniform gas parameters within the beam.
As discussed already in Sect.~\ref{sect_radex_for_map} this also leads
to very high densities.
The volume density estimates for IRS~1 and IRS~2 are much higher 
than the ones found in \citet{Koumpia2015} ($\sim$ 10 $^{6}$ cm$^{-3}$)
due to the inclusion the high--J CO transitions in the fit (13--12, 16--15).
We should rather expect that the different transitions arise from 
different layers around the infrared sources.
The fit produces a good match to all observed lines, so that we have
an accurate estimate of the total cooling through all CO lines from the
observed sources.

We use again the RADEX gas parameters to compute the C$^+$ and oxygen
column density from the line intensities given in Table~\ref{tab_co_lines}.
For IRS~1, we get column densities for C$^+$ and O of $7\times 10^{17}$~cm$^{-2}$
and $1.8\times 10^{18}$~cm$^{-2}$, respectively, for IRS~2 $5\times
10^{18}$~cm$^{-2}$~cm$^{-2}$ and $6\times 10^{18}$~cm$^{-2}$~cm$^{-2}$. 
When comparing these values with the map analysis, we notice that the
C$^+$ column density towards IRS~2 is increased by a factor two because
we use the full line width now and extract the intensity exactly
at the position of the fine-structure line peak. All other values are
basically unchanged.
At the interface, we obtain a C$^+$ column density of $5\times 10^{18}$~cm$^{-2}$
and an oxygen column density of $3\times 10^{19}$~cm$^{-2}$. We can compare 
these values with the column density that we obtain from the
total gas column density measured by CO and the upper limits of the
abundance of the two species discussed in Sect.\ref{sect_radex_for_map}.
In this way we obtain column densities of $N($C$^+)= 2\times 10^{18}$~cm$^{-2}$ 
and $N($O$) = 5\times 10^{18}$~cm$^{-2}$, i.e. much lower than computed here. 
The explanation must be
given by the PDR nature of the cloud interface. As most CO is dissociated
there, it is no longer a good measure for the total column density.
The actual column density must be at least a factor six higher when
using the gas temperature obtained from CO.

\subsection{Cooling balance}
\label{sect_cooling_balance}

The ratio of the energy emitted by the interstellar gas in the 
two main far-infrared cooling lines (\OI{} and \CII{}) relative to 
the integrated far-infrared continuum emitted by the dust can be 
used to estimate the gas-heating efficiency when assuming
that the far-infrared flux reflects the total energy of the incident
stellar radiation field by converting it to infrared-wavelengths
through absorption and re-emission in an optically thick medium
\citep[see e.g.][]{Okada2013}. As access to the \OI{} line became
possible only very recently, the \CII{}/FIR intensity ratio is
often used to measure the efficiency of the gas heating.  {\changed
This allows to address variations in the efficiency of the
photoelectric heating from dust grains because 
photoelectric heating} is the most important gas heating
process in PDRs \citep{HollenbachTielens1999}.

Although the definition of the far-infrared varies between 
different papers \citep[e.g. as FIR integrated from 42 to 122~\micron{}, 
or as TIR integrated from 3 to 1100~\micron{}; ][]{DaleHelou2002}, 
the typical range of the gas heating efficiency in Galactic PDRs, 
star-forming regions in LMC and M33 is $10^{-3}$--$10^{-2}$ both 
in case of being traced by \OI{}$+$\CII{}
\citep{Okada2013,Lebouteiller2012,Mookerjea2011,Mizutani2004} and 
traced only by \CII{} \citep{Okada2015,Lebouteiller2012,Mookerjea2011}.
As an extreme case, \citet{Vastel2001} found an efficiency of 
$10^{-4}$ in W49N, which is illuminated by an intense UV field.

\begin{figure}
   \centering
   \includegraphics[angle=90,width=7cm]{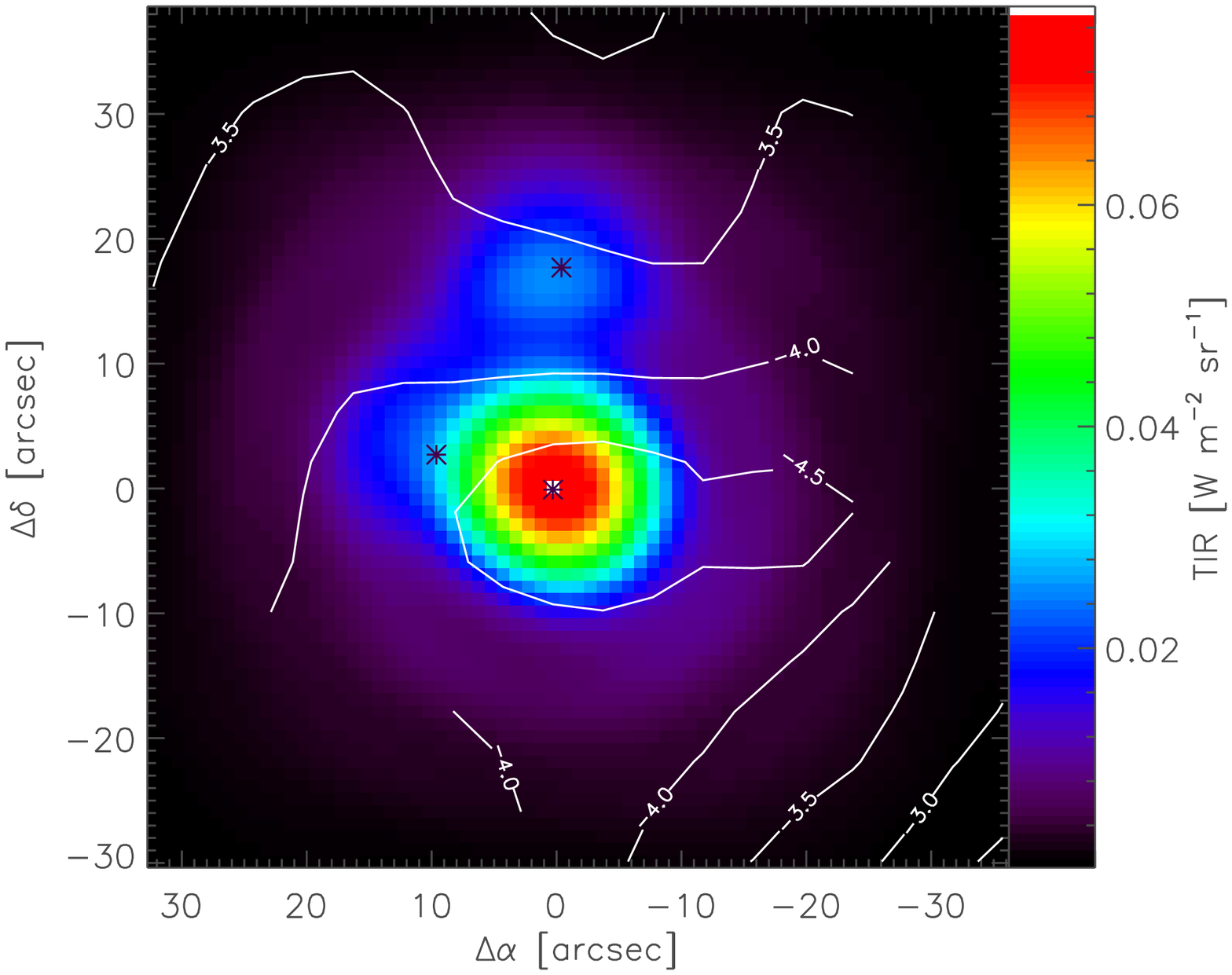}\\
   \includegraphics[angle=90,width=7cm]{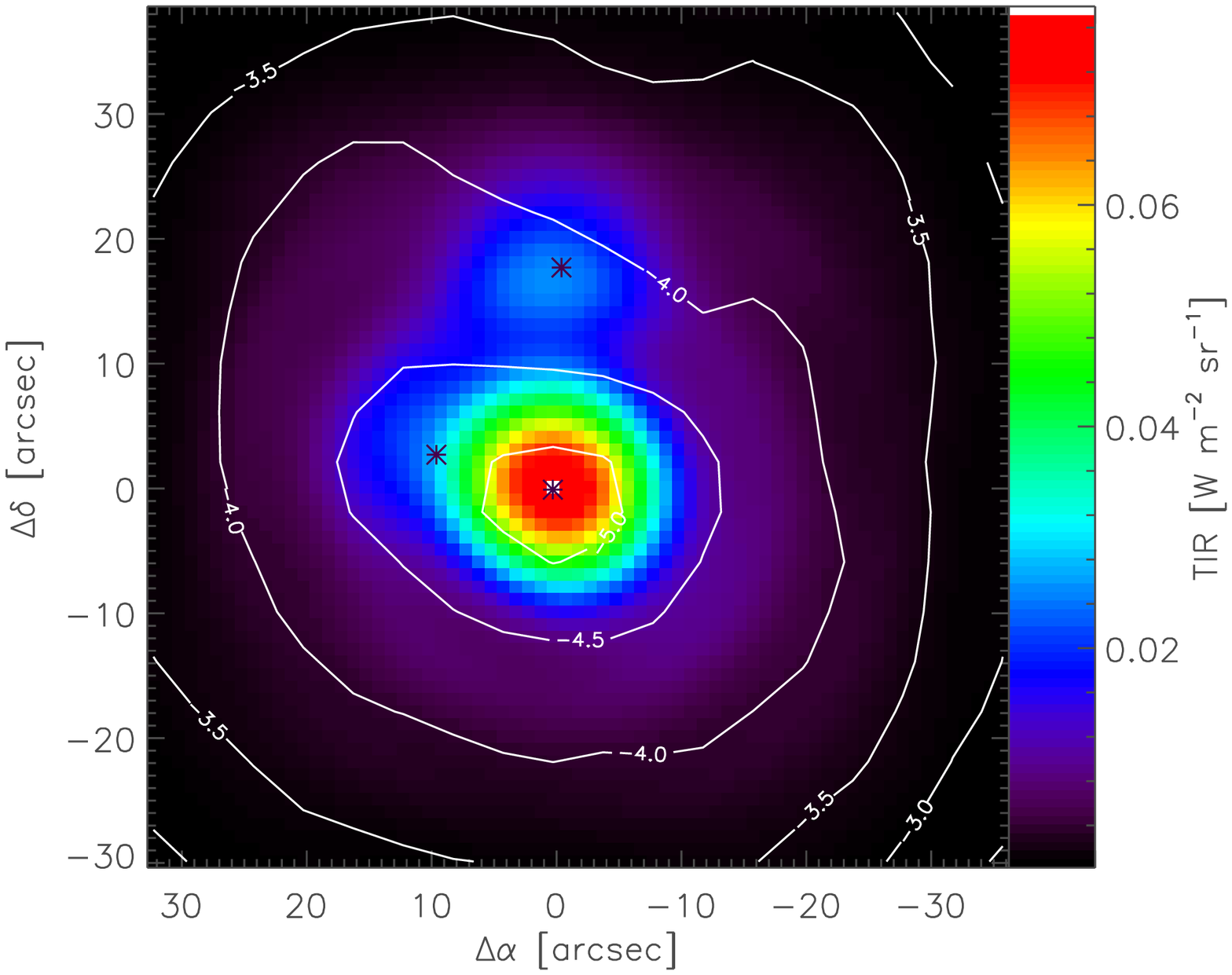}
      \caption{Comparison of the line cooling with the infrared
      continuum cooling. The background colors show the 
      total infrared flux integrated from wavelengths starting
      at 11~\micron{} \citep{Koumpia2015}. The contours give
      the decadic logarithm of the ratio of the integrated line flux 
      relative to the continuum flux. In the top panel we use
      the sum of the \OI{} and \CII{} fluxes, in the lower panel, 
      we only show the \CII{}/TIR ratio as used in many existing
      investigations.}
      \label{fig_line-fir-ratio}
\end{figure}

For the region around the embedded sources, the far-infrared
luminosity was derived from PACS and SOFIA-FORECAST observations
by \citet{Koumpia2015}. It was integrated from the observed 
intensities between 11 and 187~\micron{} further extrapolated
linearly to zero flux levels at longer wavelengths. 
Figure~\ref{fig_line-fir-ratio} compares the energy radiated in our
fine structure lines with the total infrared flux ($\lambda >
11\micron$) at the resolution of 13$''$ from the PACS continuum 
data\footnote{Our line cooling estimates for the $8.3''$ source 
at IRS~2 in Sect.~\ref{sect_irs2} slightly deviate from the numbers 
given here because of the coarser resolution applied here
compared to the direct fit of the emission profile.}
In the background we see the infrared flux from the
three embedded sources with luminosities of 10000~$L_\odot$ (IRS~1),
2000~$L_\odot$ (IRS~2), and 1300~$L_\odot$ (IRS~3), respectively
\citep{Koumpia2015}. The contours show the ratio of the sum of
the \OI{} and \CII{} fluxes relative to the total infrared (upper 
plot) and the \CII{} line only (lower plot) in logarithmic units.

The \CII{}/TIR ratio shows an almost spherically symmetric picture
around IRS~1, being dominated by the variation of the infrared 
continuum flux. The values of $10^{-3}$--$10^{-2}$, typically 
observed in other Galactic PDRs, are only seen at the 
south-western edge of the mapped area closest to the illumination 
by HD~211880. Around the embedded sources, the infrared
luminosity grows without a corresponding growth of the
fine-structure line luminosity resulting in \CII{}/TIR ratios 
below $10^{-5}$ close to IRS~1. The fine structure peak at IRS~2
only weakens this global scenario slightly, resulting in a ``dip''
in the contours next to IRS~2. 

When including the \OI{} line
in the comparison, the ratio increases by a factor three
close to the sources and the fine-structure line peak close to
IRS~2 creates a plateau in the energy ratio a about $2\times 10^{-4}$ 
in agreement to the results obtained in Sect.\ref{sect_irs2}.
At lower densities towards the map boundaries the \OI{} contribution
is negligible. 
For IRS~1, even the cooling through the lines of the CO ladder is
four times stronger than the sum of the \CII{} and \OI{}
cooling power (see. Fig.~\ref{fig_sedplot}), for IRS~2 CO sums up 
to half of the fine-structure line cooling. Including the CO
rotational ladder in the gas cooling budget raises the line-to-continuum
ratio to $1-3\times 10^{-4}$, i.e. values that are still
extremely small in Galactic context and that would be interpreted
as FIR line deficit when observed in external galaxies
\citep[e.g.][]{Luhman1998, Malhotra2001, HerreraCamus2015}. However,
none of the proposed explanations for the FIR line deficit in 
external galaxies applies to our case (see Sec.~\ref{sect_discussion}).


\begin{figure}
   \centering
   \includegraphics[angle=90,width=7cm]{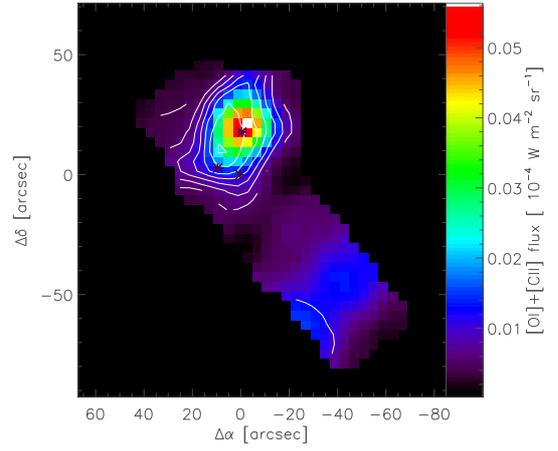}  
      \caption{Comparison of the total line cooling flux
      provided by the sum of the \OI{} and \CII{} lines (colors)
      to the \OI{}/\CII{} line flux ratio (contours from 0.4 to
      2.8 in steps of 0.4). The computation is performed in a
      common effective beam of 15$''$.
      }
      \label{fig_oi-cii-ratio}
\end{figure}

The ratio between the \OI{} and \CII{} intensities is considered as 
one of the main characteristics of PDRs. For high UV fields, the gas
temperature is always above the upper level energy of both transitions
so that the ratio between the two lines directly measures the
density \citep{Kaufman1999}. However, due to the higher energy of the
\OI{} transition it falls off quite fast at low UV fields resulting
in negligible \OI{} cooling for low densities and UV fields
\citep{Roellig2006}. 

As visible already from Fig.~\ref{fig_line-fir-ratio}, the 
relative contribution of \OI{} and \CII{} to the cooling changes
significantly over the cloud.
Figure~\ref{fig_oi-cii-ratio} shows the total line cooling budget
of the \OI{} and \CII{} lines in colors and the \OI{}/\CII{} line 
ratio in contours. We find a relatively constant ratio of 2.5--3
in the whole area of the central cluster connecting IRS~1 and IRS~2.
Here, \OI{} is dominating the gas cooling.
A value of three is representative for either very high UV fields
and low densities or low UV fields and very high densities. For the
high densities and UV fields around the central cluster, we would
however expect much higher values due to the high abundance and
emissivity of oxygen in that regime. 

When leaving the area of the central cluster, the contribution
of \OI{} to the total line cooling budget quickly drops off.
This should reflect a combination of a reduced density and 
temperature as measured by CO (Fig.~\ref{fig_RADEX_map}). While 
the density drops more quickly south of IRS~1, the temperature drops more quickly north of IRS~2. The asymmetry of the \OI{}/\CII{} ratio
plot indicates that the density is the dominating factor here.
In the bulk of the cloud and the interface, the cooling strength
of the \OI{} line can be completely neglected. \CII{} is the 
only relevant gas cooling line there.

\subsection{PDR model interpretation}
\label{sect_pdr_model}

To overcome the RADEX assumption of constant excitation conditions 
within the emitting region as an oversimplification
we switch to theoretical models for PDR structures that 
self-consistently compute the temperature and chemical structure
of PDR layers around illuminating sources, allowing us to
compare the line observations with the model predictions for 
different densities and radiation fields.

One critical input parameter is the strength of the impinging UV
radiation field. It can be computed by assuming
that in an optically thick configuration all UV photons are
eventually converted into observable infrared photons. Then we
obtain the strength of the incident radiation field in terms of
the Draine field \citep{Draine1978} as:
\begin{equation}
\chi = \frac{I\sub{TIR} \times 4\pi}{2 \times 2.7 \times 10^{-6} 
{\rm W/m^2}} 
\label{eq_uvflux}
\end{equation}
where the factor two in the denominator accounts for half of
the heating photons at wavelengths out of the UV band 
\citep[e.g.][]{Roellig2011}. Using the total infrared
flux values from Sect.~\ref{sect_cooling_balance} results in
$\chi = 5.6\times 10^4 \chi_0$ for IRS~2 and $1.2\times 10^5 \chi_0$ 
for IRS~1. For the interface, the UV field has been derived
previously to fall between $140\,\chi_0$ \citep[][]{PoelmanSpaans2005}
and $230\,\chi_0$ \citep{Timmermann1996}.

As a simple and straight-forward model, we can treat the PDRs in S~140
as plane-parallel slabs simulated by \citet{Kaufman1999}. In the model
line and continuum intensities are self-consistently computed for a
face-on configuration taking the optical depths of the lines into
account. This allows us to directly read the source parameters from 
the \CII{} line intensity, the \OI{}/\CII{} intensity ratio, 
and the (\OI{}+\CII{})/TIR ratio 
\citep[Figs. 3-6 from][]{Kaufman1999}.  As the model has only
the two parameters gas density and incident radiation field, the problem 
is over-determined allowing us to assess the consistency of the
description. Depending on the source geometry, we may, however,
introduce the inclination angle as an additional free parameter
providing a line-of-sight increase of the column densities because
the model is only computed for a face-on PDR.

This model is clearly appropriate for the external cloud
surface representing an almost edge-on PDR illuminated by HD~211880.
The main \CII{} and \OI{} emission from a plane-parallel PDR stems from
gas within $A_V \le 2$ from the cloud surface \citep{Roellig2007}.
This is also the approximate scale of chemical stratification in
the PDRs \citep{HollenbachTielens1999}. In Fig.~\ref{fig_cii-peakintens}
we can directly measure the stratification between CO 2-1 and \CII{} 
as 15$''$ or 0.05~pc. This allows to estimate the gas density as
$n\sub{H} \approx 2 \times (N\sub{H}/A_V) / 0.05\;{\rm pc} =
2.5\,10^4$~cm$^{-3}$. 

When reading
the model predictions for the given parameter combination of density
and previously fitted UV field for the interface, we obtain a 
\CII{} line intensity of $3\times 10^{-7}$~W~sr$^{-1}$~m$^{-2}$,
a \OI{}/\CII{} energy ratio of 5, and a line-to-continuum ratio of 0.02.
Unfortunately, we do not have numbers for the infrared continuum
flux at the interface position as it was outside of the region observed 
by FORECAST and PACS. We can, however, assume that the (\OI{}+\CII{})/TIR 
ratio falls above the value of $10^{-2}$ seen in the south-western edge 
of the mapped area (Fig.~\ref{fig_line-fir-ratio}).
The measured \CII{} line intensity is, however,  
$10^{-6}$~W~sr$^{-1}$~m$^{-2}$, a value that is obtained in the model
only for UV fields above $2\times 10^4 \chi_0$ and high densities. In
contrast, the measured \OI{}(63~\micron)/\CII{} energy ratio is 0.3, 
a value that is found in the model only for UV fields of about 
30~$\chi_0$ and below.

A solution can be achieved by introducing the inclination angle 
of the PDR as additional parameter and changing the assumption on
the impinging UV field. If the PDR inclination provides a geometrical
magnification of the \CII{} intensity in the optically thin case and
we assume that \OI{} is optically thick so that it is not amplified 
in the same way, the observed \CII{} intensity can be corrected 
downwards and the \OI{}/\CII{} ratio upwards. For an amplification 
factor of eight, we obtain a match of both quantities with the model
at a UV field of 60~$\chi_0$. This is lower than obtained in the 
previous PDR model fits, but still six times higher than measured
deeper in the S~140 region by \citet{Li2002}. Because our gas density
value is based on a rough estimate only, we should assume a
a factor two error bar for all numbers for a conservative approach. 
The geometrical amplification of a factor eight is obtained
by an inclination angle of 80--85~degrees relative to the face-on 
orientation. The study of the \OI{}(145~\micron{})/\OI{}(63~\micron{}) 
line ratio based on PACS data in Appx.~\ref{sect_oi_145micron} 
suggests a somewhat lower amplification factor of four. 

We can try to use the same model as an approximation for the internal 
PDR around the cluster at IRS~2 that gives rise to the observed 
\CII{} and \OI{} peak. Here, we assume that the PDR will form a 
shell around the cluster with a face-on front and back side. 
For the gas density, the value of $1.4 \times 10^{5}$~\pccm{} from
Sect.~\ref{sect_irs2} provides a lower limit. Reading the observables
from the model for this density and the UV field of $5.6\times 10^4 \chi_0$
gives a \CII{} line intensity of $1\times 10^{-6}$~W~sr$^{-1}$~m$^{-2}$,
a \OI{}/\CII{} energy ratio of 40, and a (\OI{}+\CII{})/TIR ratio
of $3\times 10^{-3}$. Comparing this to our measured values of
$1.5\times 10^{-6}$~W~sr$^{-1}$~m$^{-2}$, 5.6, and $2\times 10^{-4}$
we find a good agreement for the \CII{} line intensity when
assuming that the radiation from both surfaces adds up in the
optically moderately thick line ($\hat{\tau} = 2$, see
Sect.~\ref{sect_line_profiles}). However, the observed \OI{} line 
intensity and is about seven times lower than predicted resulting
in a line-to-continuum ratio that is also six times lower.

The much lower \OI{} intensity in the observations may stem from the 
shell configuration where we expect a decreasing \OI{} excitation
temperature towards the observer in contrast to the face-on
configuration of the PDR model where we directly observe the
illuminated side of the PDR. For \CII{} both configurations do not 
make a big difference as C$^+$ is only abundant in the narrow
$A_V < 2$ layer, but atomic oxygen is also abundant in the cool
environment at the back side of the PDR, potentially absorbing a 
large fraction of the \OI{} emission from the PDR. If we assume that
this self-absorption is as strong as a factor seven, much higher than
estimated from the line profiles in Sect.~\ref{sect_line_profiles},
we obtain a self-consistent picture for the PDR of IRS~2 in the
frame of the PDR model from \citet{Kaufman1999}. This self-absorption by a factor of seven is also consistent with the \OI{}(145~\micron{})/\OI{}(63~\micron{}) ratio observed by PACS (Appx.~\ref{sect_oi_145micron}).

By further inspecting the model output as a function of density, 
we can also constrain the maximum density of the PDR. When increasing
the density by a factor ten above our lower limit, the \CII{}
intensity drops by a factor two while the predicted \OI{} intensity
remains approximately constant. In the shell configuration this
cannot be compensated by a variable inclination angle so that we
can constrain the actual density range to densities below $10^6$~\pccm{}.
This is consistent with the central density estimate of $9\times
10^5$~\pccm{} from the dust modelling in \citet{Koumpia2015}.
With the assumption of a very strong \OI{} self-absorption, the
fine-structure line peak at IRS~2 is thus still consistent with
the standard PDR picture in a UV field of $5.6\times 10^4 \chi_0$
and a density between $1.4 \times 10^{5}$ and $1 \times 10^{6}$~\pccm{}.

However, we do not succeed any more in providing a consistent
PDR model explanation for the fine structure line deficit from 
IRS~1. The combination of the \CII{} intensity of $4\times
10^{-7}$~W~sr$^{-1}$~m$^{-2}$, the \OI{}/\CII{} energy ratio 
of 1.6, the line-to-continuum ratio of $2\times 10^{-5}$, and
the known UV field of $1.2\times 10^5 \chi_0$ provides a solution
for a gas density of $50$~\pccm{} and a factor 1.5 self-absorption
of the \OI{} 63~\micron{} line only, but this low density is in
clear contradiction to the analysis of the CO lines in 
Sect.~\ref{sect_radex_for_map} and the dust emission profile 
in \citet{Koumpia2015}.
In fact, the density around IRS~1 should be at least as high
as towards IRS~2.

\section{Discussion} 
\label{sect_discussion}

Our observational results raise two main questions:
\begin{itemize}
\item{}Why is the fine-structure line cooling from IRS~1 so weak 
compared to the continuum, with \CII/TIR ratios as low as $10^{-5}$?
\item{}Why do we see significantly more fine-structure line emission
from the weaker sources IRS~2 compared to IRS~1?
\end{itemize} 

We can discard a number of easy explanations for the lack of \CII{} 
emission from IRS~1 and the relative weakness of the \OI{} and 
\CII{} emission from IRS~2:

High resolution radio continuum observations using the VLA and MERLIN
\citep{Tofani1995, Hoare2006} showed ultracompact \HII{} regions
at the positions of IRS~1 and IRS~2. This shows that the embedded
sources produce enough UV photons with energies above 13.6~eV to
ionize hydrogen. Therefore, they should also be able to ionize
carbon (11.2~eV) and PAHs and dust grains leading to C$^+$ production
and a significant photoelectric heating of the gas. In the
same direction \citet{Malhotra2001} proposed radiation from an 
older stellar population with an overall redder spectrum to explain 
the \CII{} deficit, but in our sources we can be sure to find a 
rather very young cluster.

PACS observations of the \NIII{} 57~\micron{} line in the
frame of the Herschel key project WADI \citep{WADI} did not provide
any detection of the line in $47''\times 47''$ area centered at IRS~1.
Because of the similarity of
the second ionization potential of nitrogen and carbon, the lack
of N$^{++}$ rules out that carbon is also in the form of C$^{++}$
not contributing to the \CII{} line. Together with the non-detection
of the \OIII{} line, this rules out that a high ionization
parameter explains the lack of \CII{} and \OI{} emission.
Unfortunately, the existing
Spitzer IRS spectra do not allow to constrain the abundance of
other ionized species as they are saturated in most of the field
due to the large continuum flux.
{\changed Future observations of infrared lines might show 
whether there is a general line deficit from the \HII{} regions
that might be due to strong UV absorption by dust within
the \HII{} regions preventing any FUV photons from arriving at
the molecular gas around them.}

The intensity of the far-infrared lines may be reduced by
dust extinction for very large column densities 
\citep[see e.g.][]{Etxaluze2013}. Mainly constrained by 
450~\micron{} SCUBA observations we derived in \citet{Koumpia2015}
a total visual extinction of $A\sub{V} \approx 50$ towards
IRS~1, and of up to $A\sub{V} \approx 100$ towards IRS~2.
The dust opacities should fall between the values for
diffuse clouds \citep[82~cm$^2/g$ at 63~\micron{}, 
12~cm$^2/g$ at 158~\micron{},][]{LiDraine2001} and those for
dense clumps \citep[211~cm$^2/g$ at 63~\micron{}, 
41~cm$^2/g$ at 158~\micron{},][model 5]{OssenkopfHenning1994}
providing optical depths of $\tau\sub{63\micron}=0.1-0.25$
and $\tau\sub{158\micron}=0.01-0.05$ for IRS~1 and
$\tau\sub{63\micron}=0.2-0.5$ and $\tau\sub{158\micron}=0.03-0.2$
for IRS~2 when we assume that half of the dust is located in 
front of the line-emitting gas. This may provide a reduction 
of up to 40\,\% for the \OI{} intensity towards IRS~2, 
but only a negligible effect for all other cases, in
particular for the \CII{} line, not explaining the overall 
large deficiency.

\citet{HerreraCamus2015} proposed a redistribution of
the cooling power from \CII{} to the \OI{} 63~\micron{}
line in dense material as an explanation for the \CII{} deficit, 
but as we observed both lines, we can be sure that this
can only apply for the self-absorbed part of the \OI{}
line as all other photons are measured. We find, however,
a redistribution of cooling power to the CO rotational
lines. They can carry as much as half of the energy of
the \OI{} line, but clearly do not account for a factor 
100 difference.

Based on the spatial and spectral information in our data
we can exclude a shock-origin of our fine structure peak.
It is about 5$''$ apart from the outflow knots from IRS~1
mapped by \citet{Weigelt2002} and the fine-structure lines 
do not show the outflow wings seen e.g. in \CI{} by 
\citet{Minchin1994} or seen in our low-$J$ CO data. 
In case of outflow-excited \OI{} emission we expect a wider 
line width in \OI{} compared to CO, as observed e.g. in 
G5.89-0.39 by \citet{Leurini2015} and not the opposite 
behavior. Moreover, the \OI{} line has a slightly blue-skewed
profile that is rather representative for infall than for 
outflow.

In the frame of PDRs, low line-to-continuum cooling ratios
are explained by {\changed either very high or very low} charging 
parameters $\chi/n$ {\changed  \citep[e.g.][]{Kaufman1999}. 
High $\chi/n$ values lead to} positively charged dust grains 
and PAHs, lowering the photoelectric efficiency . They
are obtained for very high UV fields at low densities.
{\changed In the opposite case all material remains cold
and no C$^+$ is produced so that most energy is emitted in
CO lines.} However, our CO and dust observations exclude low densities
around the infrared sources (Sect.~\ref{sect_RADEX_for_sources}),
{\changed the measured \CII{} intensity excludes the cold-gas
solution, and the UV fields are} constrained by the measured infrared 
luminosity.

Alternatively, the photoelectric efficiency can be reduced by
the destruction of PAHs since PAHs are the dominant carrier 
of the photoelectric heating \citep{BakesTielens1994}. As 
mentioned above, the saturation of the IRS spectra around 
infrared sources prevent us to examine the PAH spectra directly, 
but when we compare the relative strength of the PAH 11.3\micron{} 
feature against the underlying continuum southwest of IRS~1, 
we find some reduction in the area closer area to IRS~1 
(20--30$''$ from IRS~1) compared to 30--50$''$ from IRS~1.
However, the spectra show still an infrared excess from
very small dust grains indicating that the reduction in
the PAH abundance can only weakly reduce the overall surface
for the photoelectric effect.

The relative weakness of the fine structure lines and the low
\OI{}/\CII{} ratio must be explained by deviations of the 
conditions in S~140 from that ``standard PDR''. For the \OI{}
line the low intensity can be explained by a scenario with
a decreasing \OI{} excitation temperature towards the observer,
resulting from a decreasing gas temperature and/or density. Due to
the high column density of atomic oxygen at low temperatures,
this geometry can provide a strong self-absorption, explaining
the factor seven reduction of the \OI{} line intensity observed
towards IRS~2. However, this mechanism can hardly be responsible
for the same kind of reduction in the \CII{} line intensity towards
IRS~1. As C$^+$ quickly recombines in a dense gas at low
temperatures, finally forming CO, we do not expect large
quantities of cold C$^+$ in the line of sight that can 
produce the same kind of self-absorption for the \CII{} fine-structure
line. A very special configuration would be needed that
allows for a significant amount of cold low-density C$^+$
gas in a clumpy medium behind a layer of dense and warm
gas facing the \HII{} region around the embedded cluster.

Beyond the relative weakness of the fine-structure lines 
relative to the continuum flux, we find puzzling combinations 
of detections:
\begin{enumerate}
\item{}IRS~2, with 2000~$L_\odot$ shows strong \OI{}, \CII{},
and CO (and isotopologues) emission at about -6.5~\kms{} and
has associated ultracompact \HII{} regions.
\item{}IRS~1, with 10000~$L_\odot$ shows only weak \CII{} and
\OI{} emission at the main velocity of the cloud of -8~\kms{}.
The CO lines have a somewhat smaller intensity than on IRS~2
and the optically thin lines are centered at the IRS~2 velocity, 
showing only an emission wing at -8~kms{} indicating also a 
relative lack of CO emission from IRS~1. An ultracompact \HII{} region and
an outflow were detected.
\item{}IRS~3, with 1300~$L_\odot$ shows \OI{}, but hardly any 
\CII{}, CO is comparable to IRS~1, but spatially not separated
in our beam. The source, however shows no associated \HII{} region.
\end{enumerate}


\section{Summary}

Our GREAT observations of S~140 showed a pronounced peak of the 
emission of the \OI{} 63~\micron{} and \CII{} lines 20$''$ north 
of IRS~1, the main embedded source dominating the infrared field
in the whole region. The fine-structure-line peak can be
associated with a weaker infrared source, IRS~2, and the GREAT
observations resolve the size of the emitting region to have
a diameter of 0.03~pc. The velocity of the gas at that position
is offset by about 1.5~\kms{} from the bulk of the molecular
material in S~140. The gas density of the emitting gas
must be at least $1.4\times 10^5$~\pccm{} to allow for the observed
strength of the \CII{} line. In spite of its absolute strength,
the \OI{} 63~\micron{} line, is however, relatively weak when
compared with \CII{} and the infrared continuum. This can only be
explained by a strong self-absorption in the foreground medium
with a large optical depth. For \CII{} we measured an optical 
depth of $\hat{\tau}\approx 2$. 

From IRS~1 we find only relatively weak \CII{} and almost no
\OI{} emission in spite of its high luminosity of $10^4 L_\odot$
and a know ultracompact \HI{} region. The main heating source of 
dust and large scale-gas in S~140 is not responsible for the
\CII{} and \OI{} peak, but shows an extreme fine-structure line
deficit. While for IRS~2, the low value line of the (\CII{}+\OI{})/TIR
ratio, also matching the criterion for a line deficit, can be explained
by the strong \OI{} self absorption, this explanation does not work
for IRS~1, and in particular not for the low \CII{} intensity observed
there. Solving the puzzle of the extremely low fine-structure
in S~140 may be the puzzle to a general explanation of the
fine-structure line deficit in galaxies with high infrared luminosities.

At the cloud surface of S~140 we find a classical PDR, showing the
expected stratification in the line tracers and producing strong 
\CII{} emission. The \CII{} line is narrower there, having an
optical depth of $\hat{\tau}\approx 3$. The combination of the 
observed stratification and line intensities of \CII{} and \OI{} 
can be explained by a plane-parallel PDR model with a density of
$2.5\,10^4$~cm$^{-3}$ if we assume that the surface is tilted
by an inclination angle of 80--85~degrees relative to the face-on 
orientation and illuminated by a a UV field of 60~$\chi_0$, a value
that is lower than obtained in the previous PDR model fits in the 
literature.

\begin{acknowledgements}
      We thank Ed Chambers, Denise Riquelme, and Christoph Buchbender
      for help with the data calibration and Paul Harvey for help
      with the continuum data and many useful discussions.
      This work was supported by the German
      \emph{Deut\-sche For\-schungs\-ge\-mein\-schaft, DFG\/} project
      number SFB 956, C1.
      
      The work presented here is based on observations made with the NASA/DLR
Stratospheric Observatory for Infrared Astronomy. SOFIA Science Mission Operations
are conducted jointly by the Universities Space Research Association,
Inc., under NASA contract NAS2-97001, and the Deutsches SOFIA Institut under
DLR contract 50 OK 0901. We greatfully acknowledge the support by the
observatory staff.
      
       HIFI has been designed and built by a consortium of institutes and university departments from across Europe, Canada and the United States under the leadership of SRON Netherlands Institute for Space Research, Groningen, The Netherlands and with major contributions from Germany, France and the US. Consortium members are: Canada: CSA, U.Waterloo; France: CESR, LAB, LERMA, IRAM; Germany: KOSMA, MPIfR, MPS; Ireland, NUI Maynooth; Italy: ASI, IFSI-INAF, Osservatorio Astrofisico di Arcetri-INAF; Netherlands: SRON, TUD; Poland: CAMK, CBK; Spain: Observatorio Astronómico Nacional (IGN), Centro de Astrobiología (CSIC-INTA). Sweden: Chalmers University of Technology - MC2, RSS \& GARD; Onsala Space Observatory; Swedish National Space Board, Stockholm University - Stockholm Observatory; Switzerland: ETH Zurich, FHNW; USA: Caltech, JPL, NHSC.

PACS has been developed by a consortium of institutes led by MPE (Germany) and including UVIE (Austria); KU Leuven, CSL, IMEC (Belgium); CEA, LAM (France); MPIA (Germany); INAF-IFSI/OAA/OAP/OAT, LENS, SISSA (Italy); IAC (Spain). This development has been supported by the funding agencies BMVIT (Austria), ESA-PRODEX (Belgium), CEA/CNES (France), DLR (Germany), ASI/INAF (Italy), and CICYT/MCYT (Spain).
\end{acknowledgements}


\bibliographystyle{aa}

\Online

\begin{appendix}

\section{Comparison to Herschel data}
\label{sect_appx_herschel}

In the frame of the WADI key project \citep{WADI}, S140 was observed
with the HIFI and PACS instrument, also covering the \CII{} and \OI{}
lines\footnote{Herschel obsids: PACS:
1342222255 (IRS~1 \OI{} ~63\micron), 1342222256 (IRS~1 68--208~\micron),
1342222257 (Interface \OI{} 63~\micron), 1342222258 (Interface \OI{}
145~\micron), 1342222259 (Interface \CII{} 158~\micron), 
HIFI: 134290781 (\CII{})}. 
Part of the results were used e.g. in the papers of \citet{Dedes2010} and
\citet{Koumpia2015}. Due to the limited observing time in the key project
only a small part of the source was covered. In the overlapping part
we use the data here to check the consistency with our new, more extended
observations with GREAT.

   \begin{figure}
   \centering
   \includegraphics[angle=0,width=\hsize]{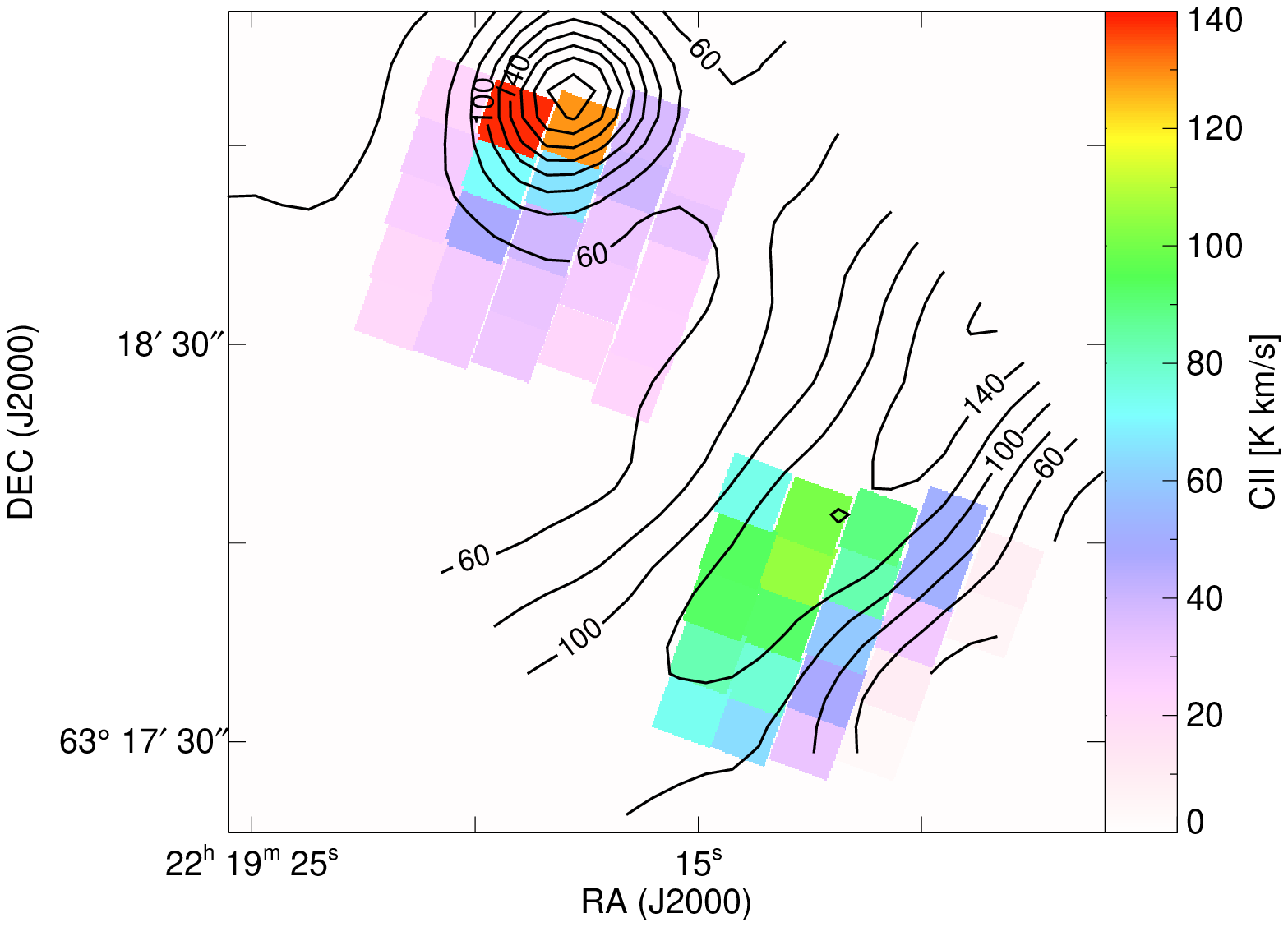}\\
   \includegraphics[angle=0,width=\hsize]{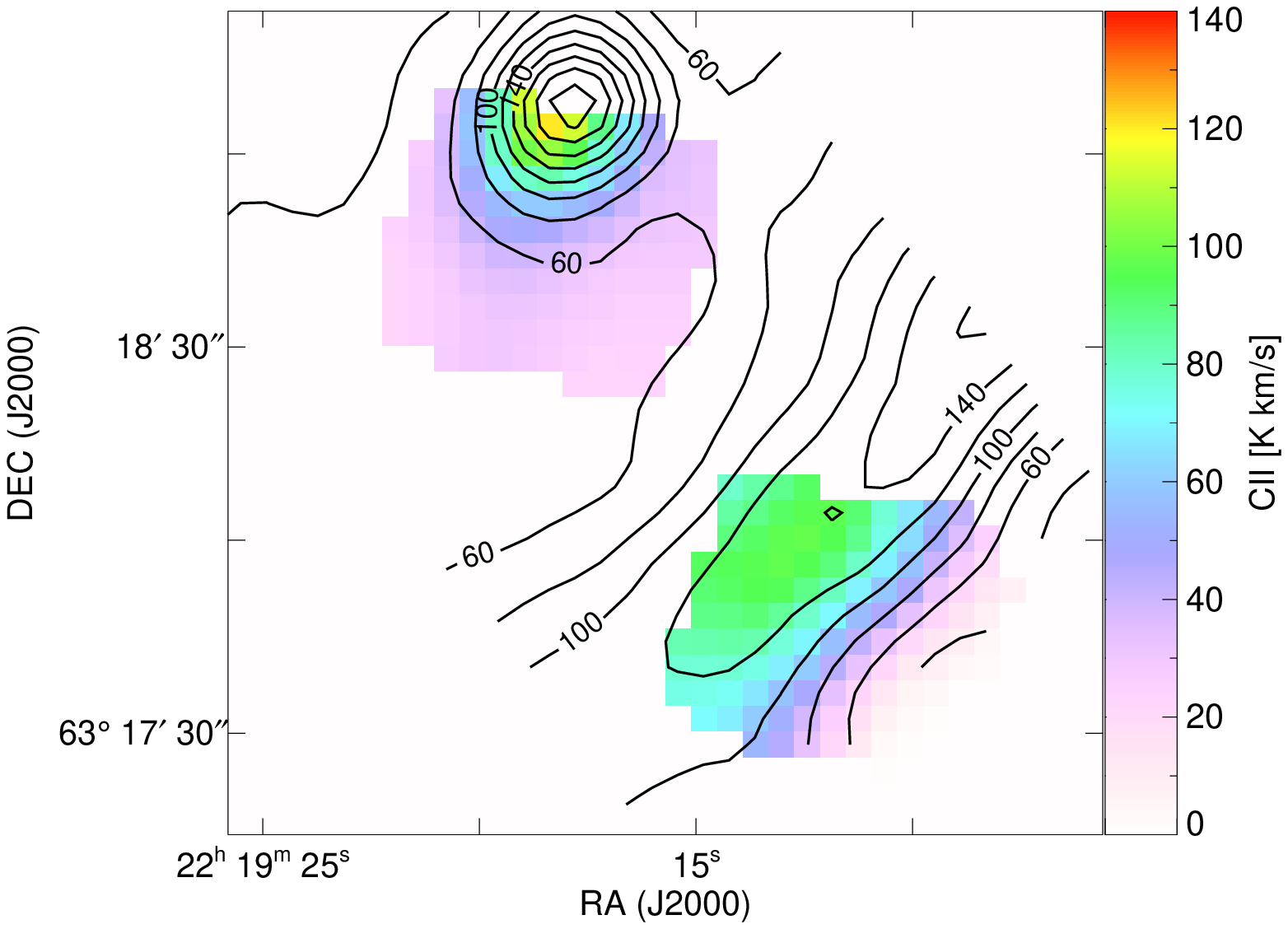}
      \caption{Integrated map of \CII{} obtained in the GREAT observations
		(contours) overlaid by colors of the \CII{} line
		intensity from the two PACS footprints. The top image shows
		the intensities in the original PACS footprints. In the bottom
		version this is resampled to the positions of the GREAT map
		with an effective resolution of 17.2$''$ beam.}
         \label{fig_cii-pacs}
   \end{figure}

\subsection{PACS spectroscopy}
\label{sect_pacs_comparison}

The \OI{} and \CII{} lines were observed earlier in S~140 through
the PACS instrument onboard Herschel in the frame of a spectral
scan at two individual pointing, one centered at IRS~1 and the other
one close to the cloud surface towards HD~211880. The PACS
footprint is not fully-sampled so that it may miss part of the flux
between the pixels, but if we assume a source size of 5.5$''$ or above
that missing flux should be small so that we can use the data
to check the consistency with the GREAT observations.
The spectral resolution of PACS was insufficient
to resolve the line, so that we can only compare integrated
intensities.

   \begin{figure}
   \centering
   \includegraphics[angle=0,width=\hsize]{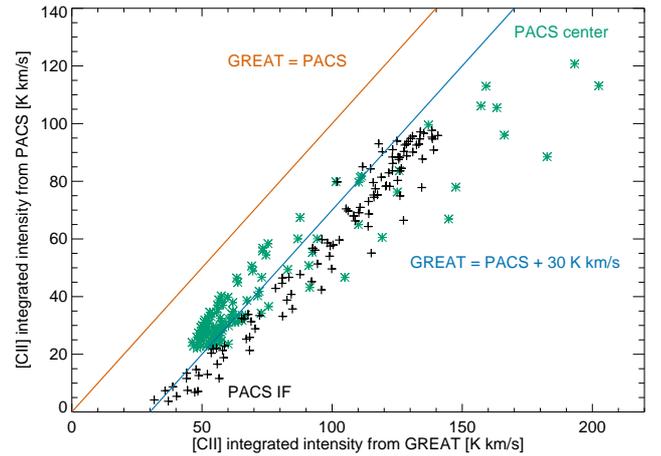}
      \caption{Measured intensities in the different pixels from
	Fig.~\ref{fig_cii-pacs}b as seen by GREAT and PACS. Green
	asterisks represent the footprint around IRS~1, black plus signs
	the footprint at the outer interface. The brown line shows
	the identity, the blue line represent an identity with an
    offset of 30~K~\kms{}. }
         \label{fig_cii-pacs_correlation}
   \end{figure}

Figure~\ref{fig_cii-pacs} compares our new intensity map
with the PACS data for the two footprints. We see that the spatial
structure of the emission peaks around IRS~2 and the interface
is consistent between PACS and GREAT data. The PACS data just miss the
global emission peak, but they peak at the position closest to the
true maximum and they also trace the structure of the extended emission
at the outer cloud interface. When comparing absolute intensities,
however, we find systematically somewhat lower intensities in the
PACS data compared to GREAT data.

   \begin{figure}
   \centering
   \includegraphics[angle=0,width=\hsize]{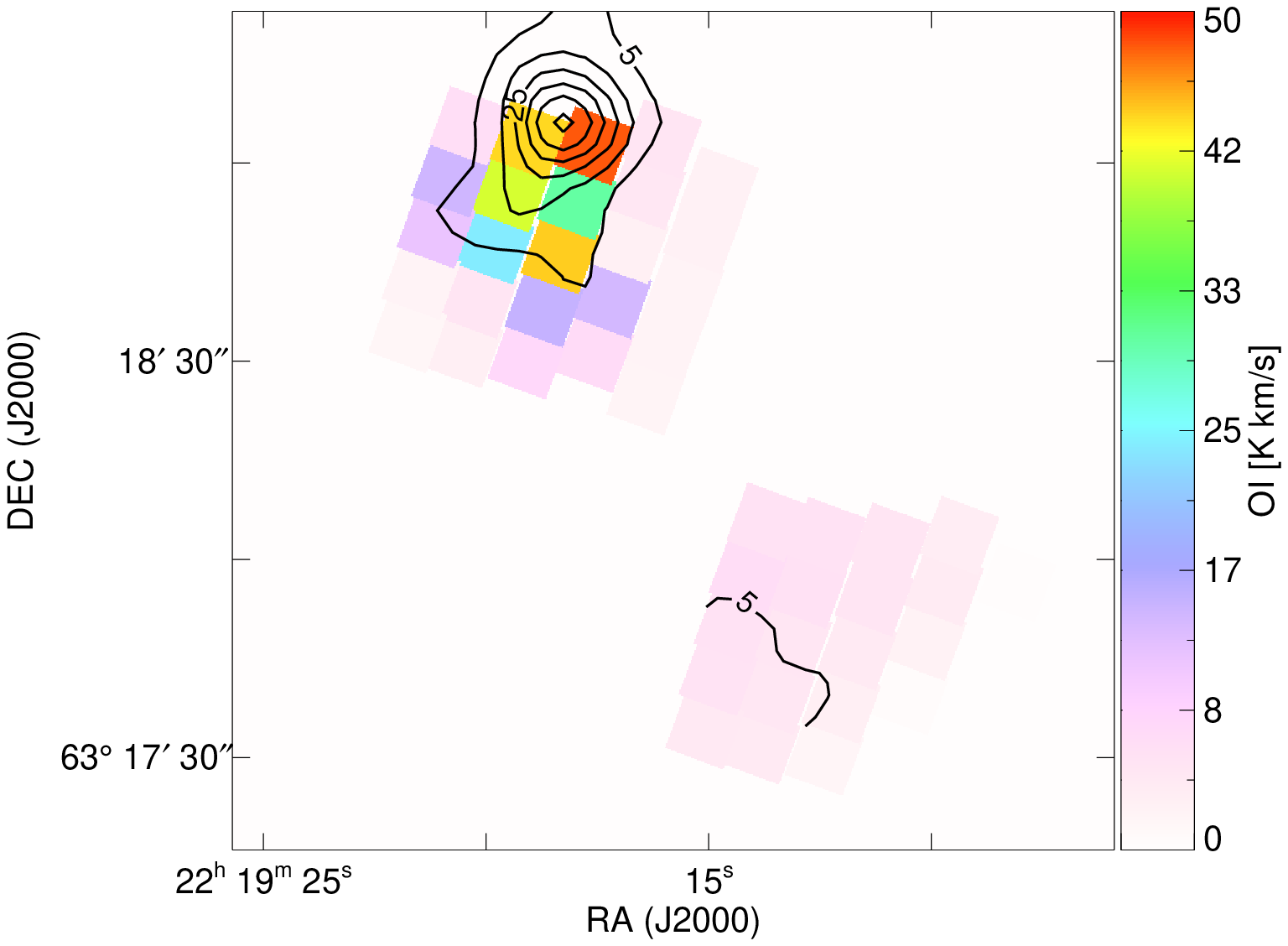}\\
   \includegraphics[angle=0,width=\hsize]{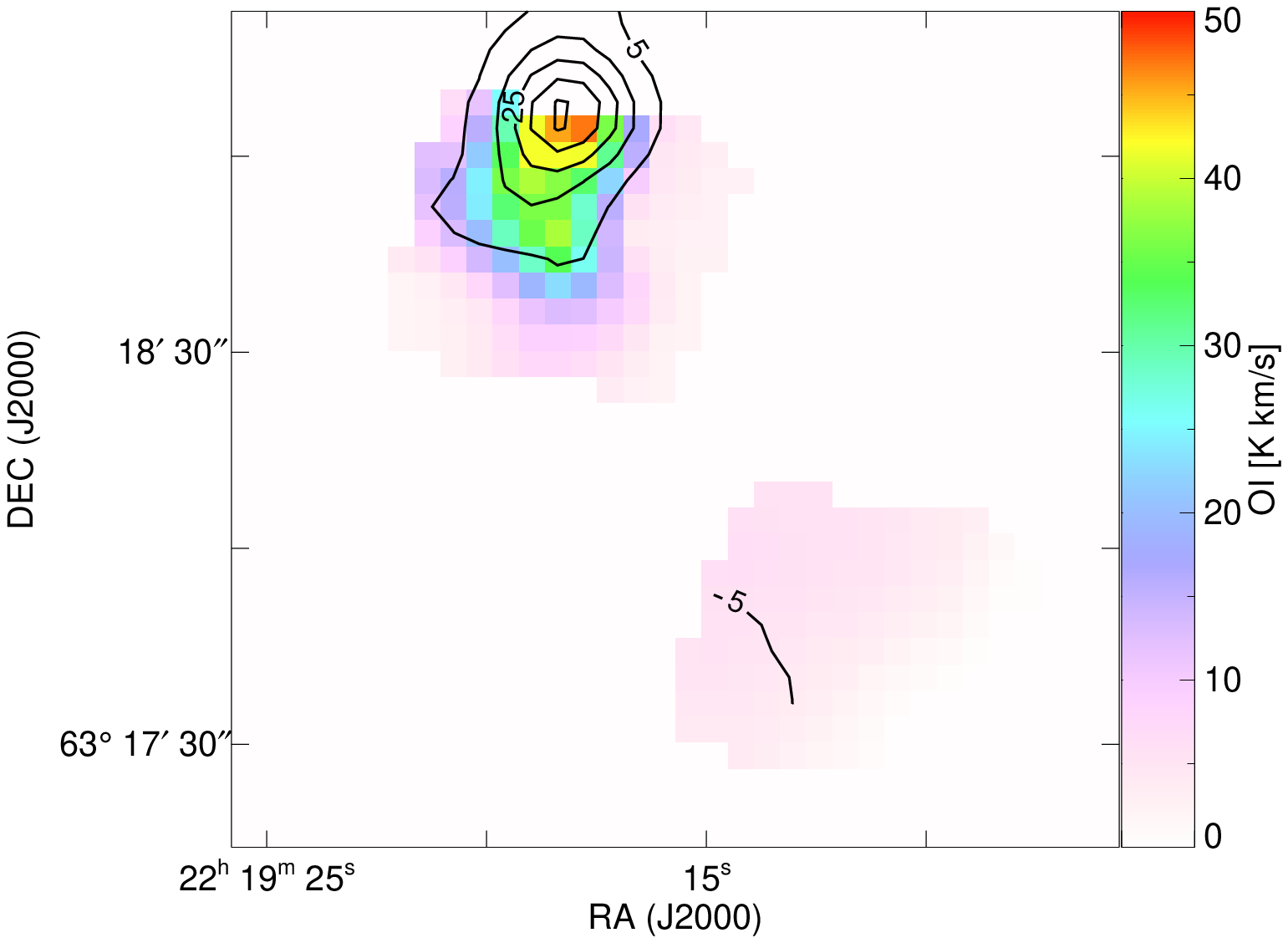}
      \caption{Integrated map of \OI{} obtained in the GREAT observations
		(contours) overlaid colors of the \OI{} line
		intensity from the two PACS footprints. The top image shows
		the intensities in the original PACS footprints with GREAT data
        convolved to 10$''$ resolution. In the bottom
		version both maps are resampled to the positions of the GREAT map
		for the effective resolution of a circular 12$''$ beam.}
         \label{fig_oi-pacs}
   \end{figure}

To understand whether this is a calibration problem, we computed the
pixel-to-pixel statistics in the resampled map. This is shown in 
Fig.~\ref{fig_cii-pacs_correlation}. We find a uniform behavior between
both footprints; the PACS intensity is offset from the identity by about
30~K~\kms{}. This means that we do not have a calibration difference,
but a difference in the absolute level. This can be naturally explained
if there is a 30~K~\kms{} contamination of the OFF position used as 
the reference in the PACS observations. The GREAT observations use a
well controlled OFF position 6.5$'$ south-west of the molecular cloud, 
while the chopped PACS observations involve a reference that falls
6$'$ into the molecular cloud. Some \CII{} emission from that region is
quite likely and would explain the constant absolute offset between the
GREAT and PACS intensities.

   \begin{figure}
   \centering
   \includegraphics[angle=0,width=\hsize]{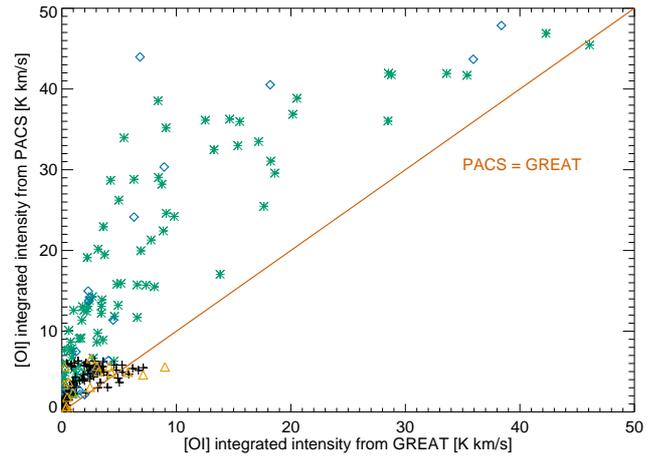}
      \caption{Measured intensities in the different pixels of
	Fig.~\ref{fig_oi-pacs} as seen by GREAT and PACS. Green
	asterisks represent the footprint around IRS~1 and black plus 
    signs the footprint at the outer interface when both data
    sets are resampled to a common 12$''$ resolution
    (Fig.~\ref{fig_oi-pacs}b). 
    Blue diamonds and orange triangles are obtained when we resample 
    instead the GREAT data to the 25 pixels of the PACS array using 
    the 9$''$ PSF of PACS for the IRS~1 region and the interface
    respectively (Fig.~\ref{fig_oi-pacs}a). 
    The brown line shows the identity. }
         \label{fig_oi-pacs_correlation}
   \end{figure}

As the \OI{} emission is much less extended, we expect no OFF
contamination in the PACS data for this line. Figure~\ref{fig_oi-pacs} 
compares the spatial distribution of the \OI{} integrated line
emission measured by PACS with our GREAT observations. The spatial distributions of the emission seen through GREAT and PACS agree relatively well, having the peak again around IRS2. But for the individual intensities we
find some significant deviations. Fig.~\ref{fig_oi-pacs_correlation} 
compares PACS and GREAT intensities from two different ways of resampling. One is shown in the lower panel of Fig.~\ref{fig_oi-pacs}, namely both maps are resampled to a common 4$''$ grid with 12$''$ resolution (green asterisk and black plus sings in Fig.~\ref{fig_oi-pacs_correlation}). Since the PACS map is not fully-sampled, we extracted the flux based on the geometrical overlap. In the second method, we used the integrated intensity of the original $5\times 5$ PACS spaxels, and resampled the GREAT data to each spaxel position (blue diamonds and orange triangles in Fig.~\ref{fig_oi-pacs}. In both ways the resulting trend is the same. There is a good match for the
map around the interface and for the highest intensities close to 
IRS~2 at the northern boundary of the PACS field. However, across
the PACS array centered at IRS~1, the intensities seem to by
systematically too high. Off-contamination on the GREAT side can
be excluded as it would show up as a constant offset like seen in 
the \CII{} case. Moreover, we used the same reference position as 
for \CII{} and \OI{} is rather less extended than \CII{}
due to the higher critical density.
On can see in Fig.~\ref{fig_oi-pacs_correlation} that the different 
ways of resampling the data do not make any significant difference. 
The PACS observer's manual (v2.3) does not provide any direct explanation.
63\micron{} is not listed as a ghost wavelength from other strong 
emission lines and it is not in the leakage wavelength range.
We can only speculate that this may be a stray-light problem where
the strong \OI{} emission just at the edge of the PACS footprint
is partially picked up in the adjacent spaxels. Unfortunately, there
are no calibration data available to verify that situation as the
calibration observations were not set up to just miss the target.
Hence, we cannot provide a quantitative explanation for the mismatch.

\subsection{HIFI cut}

   \begin{figure*}
   \centering
   \includegraphics[angle=90,width=\hsize]{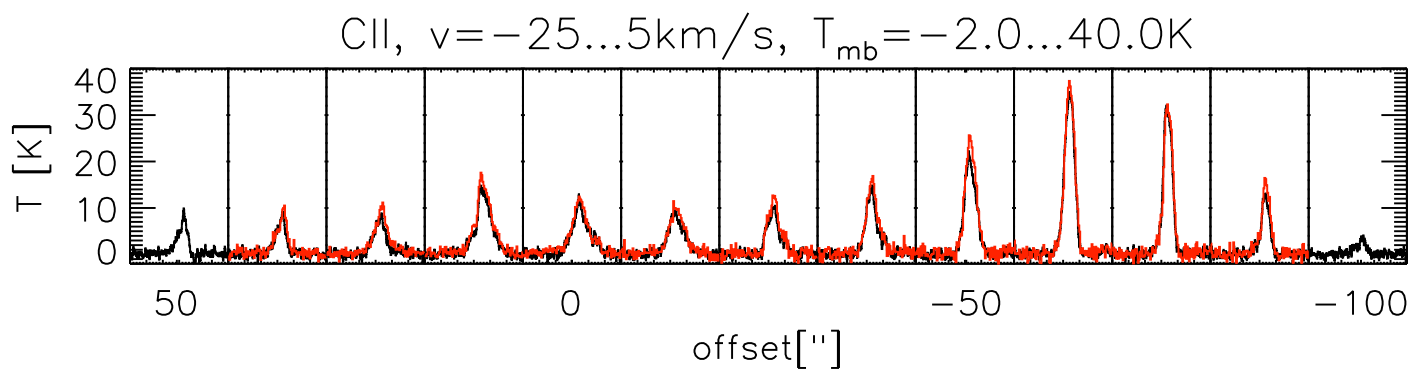}
      \caption{\CII{} spectra taken with HIFI (black) and GREAT (red)
		along a 37 degrees cut through IRS~1 and the cloud surface.
		The HIFI cut extends deeper into the cloud and only
		every second spectrum is plotted. The corresponding GREAT
		spectra were averaged in a 6$''$ circle around the HIFI
		position.}
         \label{fig_cii-hifi}
   \end{figure*}
   
The \CII{} line was also observed by HIFI \citep{Dedes2010}
in a single strip going through IRS~1 and the surface also oriented
in the same angle of 37 degrees east of north as our GREAT map.
In Fig.~\ref{fig_cii-hifi} we compare individual HIFI spectra along that cut at
a half-sampled spacing with the corresponding GREAT spectra obtained
by averaging all data in a  6$''$ circle around the HIFI positions.

We find a generally very good match between HIFI and GREAT spectra,
in particular the line shapes agree in all details, but at some
positions the GREAT peak intensities are larger by a few up to 15\,\%.
First of all this confirms the good relative calibration and pointing
accuracy of both instruments/telescopes. The higher intensity of GREAT
at an offset of $-50''$ and $+12''$ could be due to the somewhat
larger telescope beam of SOFIA leading to more intensity pickup from
the surface at $-50''$ and from the newly detected intensity peak
north of IRS~1 at $+12''$.

\section{The \OI{} 145~\micron/63~\micron{} ratio}
\label{sect_oi_145micron}

\begin{figure}
   \centering
   \includegraphics[angle=0,width=\hsize]{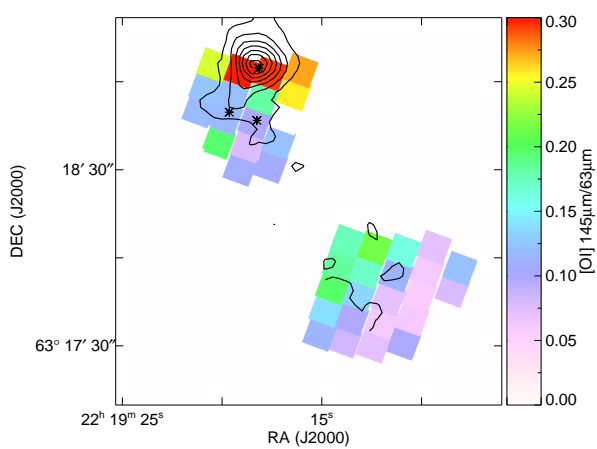}
      \caption{Observed ratio of the \OI{} 145~\micron{} to
      \OI{} 63~\micron{} line integrated intensity seen by PACS 
      (colors) overlaid on the contours of the \OI{}
      63~\micron{} emission measured by GREAT.}
      \label{fig_oi-ratio}
\end{figure}

In spite of the possible stray-light issue of the PACS \OI{}
map (see Appx.~\ref{sect_pacs_comparison}), we can also try to
use the \OI{}(145~\micron{})/\OI{}(63~\micron{}) line ratio to
constrain the parameters of the PDRs in S~140, assuming that 
the ratio between both lines 
should be more robust than the simple integrated line intensity.


Because the excitation energy of the 145~\micron{} line of 325~K 
lies above that of the 63~\micron{} line (228~K) the 
\OI{}(145~\micron{})/\OI{}(63~\micron{}) measures temperature 
(or gas heating through the UV field) in the temperature range of 
about 300~K. In many cases, however, this ratio may be dominated
by the optical depth in the 63~\micron{} line
\citep{Kaufman1999,Malhotra2001}. A low \OI{}(145~\micron{})/\OI{}(63~\micron{}) compared to the PDR model prediction is observed in many PDRs, and the foreground absorption, opacity effect of the \OI{} 63~\micron, and the cloud geometry are proposed as possible explanations  \citep{Liseau1999,Caux1999,Giannini2000,Okada2003}. \citet{Liseau2006} provided the theoretical estimates of this ratio for different physical conditions. In case of the optically thin emission, the ratio over 0.1 can be achieved only when the gas temperature is $>500$~K and the collision partner is H$_2$, which is unlikely. For optically thick case, their calculation with the density of $3\times 10^4$~cm$^{-3}$ shows that the column density of $N$(H)$>10^{24}$~cm$^{-2}$ would be required to make the \OI{}(145~\micron{})/\OI{}(63~\micron{}) ratio above 0.1.

Figure~\ref{fig_oi-ratio} shows the integrated intensity ratio of 
\OI{}(145~\micron{})/\OI{}(63~\micron{}) from the two PACS footprints 
centered at IRS~1 and at the interface observed in the frame of 
the Herschel key project WADI \citep{WADI}. 
For the interface parameters discussed above, the face-on PDR model 
produces a \OI{}(145~\micron{})/\OI{}(63~\micron{}) ratio of 0.04, 
whereas the observed value is 0.17, i.e. about four times higher.
If we interpret it through the inclination of the PDR in the same 
way as for the \CII{} line, the ratio asks for a geometrical 
amplification of the \OI{} 145~\micron{} line by a factor four,
tow times lower than inferred from the \OI{}(63~\micron)/\CII{} 
ratio in Sect.~\ref{sect_pdr_model}.

At IRS~2, \OI{}(145~\micron{})/\OI{}(63~\micron{}) from PACS observations is 0.3. For the PDR with a UV field of $5.6\times 10^4\chi_0$ and a density between $1.4\times 10^5$ and $10^6$~cm$^{-3}$ (Sect.~\ref{sect_pdr_model}), the predicted ratio is 0.03--0.045. Since the absorption by the cold gas at the back side of the PDR should affect only the 63\micron{} transition, a factor 7-10 of self-absorption is required to be consistent with the model predicted \OI{}(145~\micron{})/\OI{}(63~\micron{}.

At IRS~1, the observed \OI{}(145~\micron{})/\OI{}(63~\micron{}) is 0.1, which suggests the PDR gas with a density of $10^3$--$10^4$~cm$^{-3}$ and a UV field of $>10^6$. But as discussed in Sect.~\ref{sect_pdr_model}, there is no consistent model for IRS~1.


\end{appendix}

\end{document}